\documentclass[fleqn,10pt]{olplainarticle}

\usepackage[numbers,sort&compress]{natbib}
\usepackage{float}
\usepackage[utf8]{inputenc} 
\usepackage[T1]{fontenc} 
\usepackage{hyperref} 
\usepackage{url} 
\usepackage{booktabs} 

\usepackage{tabularx} 
\usepackage{amssymb} 

\usepackage{amsfonts} 
\usepackage{nicefrac} 
\usepackage{microtype} 
\usepackage{lipsum}
\usepackage{graphicx}
\usepackage{amsmath}
\usepackage{multirow}
\usepackage{xcolor}
\usepackage{colortbl}
\graphicspath{ {./images/} }
\usepackage{tikz}
\usepackage{tcolorbox}
\usetikzlibrary{positioning, arrows.meta, shapes.geometric}

\definecolor{best}{RGB}{144, 238, 144}
\definecolor{worst}{RGB}{255,230,230}

\title{Popularity Trajectories and Mixing Patterns of Urban Venues in Two European Cities}

\author[1, 3]{Wang Ngai Yeung}
\author[2]{Lakmal Meegahapola}
\author[2, 4]{Daniele Quercia}
\author[1, 5]{Riccardo Di Clemente}
\affil[1]{Complex Connections Lab, Network Science Institute, Northeastern University, London, United Kingdom}
\affil[2]{Nokia Bell Labs, Cambridge, United Kingdom}
\affil[3]{Public Tech Media Lab, University of Wisconsin-Madison, Madison, WI, United States}
\affil[4]{Politecnico di Torino, Turin, Italy}
\affil[5]{ISI Foundation, Turin, Italy}


\begin{abstract}
Urban venues such as restaurants, shops, bars and parks host the daily life of a city, and their collective rise and fall signals how vibrant a neighbourhood is. The usual assumption is that a venue's fortunes follow its neighbourhood's, so that how wealthy an area is and how many people live there will decide a venue's performance. Yet whether there is a common performance trajectory shared by venues over time is underexplored. In this study, we use 5.5 years of Foursquare data for about $10^4$ venues in London and Barcelona. We found six distinguishable shapes recur in both cities, that are (1) Fall, (2) Rise, (3) Rise \& Fall, (4) Steady, (5) Ephemeral (an early hit that loses popularity fast) and (6) Underdogs (a slow starter that gains popularity late). However, no recovery shape appears, that is neither city yields a U-shaped fall-and-rise group. Additionally, we found that deprivation and population density are not indicative which shape a venue follows. We also discovered that how early a shape can be recognized depends on when its distinctive segment occurs, not a shape's complexity. Proximity between venues, however, does predict how well-performing a venue is. We uncovered a mixing pattern where venues of the same shape tend to locate together more often than chance within roughly $200$--$600$\,m. Finally, our qualitative reading of two contrasting districts, one rising in London and one falling in Barcelona, show that comparable socioeconomic conditions but divergent gentrification histories can generate vastly different popularity trajectories. All in all, our results altogether point at that urban vibrancy emerges at the block- and street- level, and social and historical contexts matter in terms of the characteristics of such patterns. 
\end{abstract}

\begin{document}
\maketitle
\section*{Introduction}\label{sec:introduction}

Two cafes in the same neighbourhood can end up on opposite directions, that is one builds a following over a decade while the other, on the next street, quietly empties out. Sometimes it is hard to explain why, because both venues are located within the same neighbourhood. Urban venues such as restaurants, bars, shops and parks, where a city's everyday social and economic life takes place, are usually explained by the neighbourhood around them. They provide the shared spaces in which residents meet and communities form, and a city's long-term vitality is widely known to depend on its neighbourhoods' vibrancy, a phenomenon that has been studied with both qualitative methods and large-scale data analyses \cite{jacobs1961death, mouratidis2020built, yiu2026vibrancy, chen2024achieving, Tu2020}. Yet these venues change over time: they open, flourish, and potentially close, and the health of an urban space can bound up with a series of successes and failures of urban venues \cite{Parsa2005, psyllidis2022points}. Disentangling how venues gain, maintain, and lose popularity over time is therefore vital to business owners, urban planners, and the citizens who live in their vicinity.

Whether a business will thrive in the future is generally very hard to predict in advance, and restaurants and bars are among the hardest cases \cite{Parsa2005}. Being able to predict a business's trajectory early offers its owners better insights for strategic planning and timely correction to avert failure. Despite prior work on the short and long-term success of individuals across contexts and markets, our understanding of how performance evolves over time for urban venues remains relatively scarce. Part of the difficulty is that success is that it cannot be measured merely by raw performance. So, what separates a successful venue from the rest is how the community (e.g., audience, customers) responds to it relative to its competitors \cite{Yucesoy2016}. Throughout this paper we call that relative response a venue's popularity. The proliferation of location-based and transactional data has made measuring such ``relative success'' more accessible at fine spatio-temporal resolution. Credit-card transactions uncover the lifestyles and consumption patterns of urban populations \cite{di2018sequences} and GPS and mobile-phone traces reveal the temporal rhythms of streets and neighbourhoods \cite{su2022rhythm}. In our case, crowdsourced platforms such as Yelp and Foursquare allow us access to signals of popularity, such as reviews, likes and check-ins of every venue, such that a venue's popularity can be compared with another's at the same moment. Crucially, such data are longitudinal, offering a granular digital trace of a venue's popularity over time.

Several factors are thought to shape trajectories of venue popularity. The first set of factors is internal to the venue, such as product quality, pricing and branding. The second is what other people say about it --- online reviews have been shown to act as public word-of-mouth, shaping business outcomes from revenue to how often a venue sells out \cite{Anderson2012, luca2016reviews}. The third is the surrounding urban environment, which matters for two reasons. Similar businesses clustered together can draw more custom than they would alone, an effect called agglomeration \cite{glaeser2010agglomeration}. Secondly, the surrounding population, including its income, its density, and other existing venues, also affects a venue's performance overall, which is known as a neighbourhood effect \cite{Yuan2021}. The most studied case is gentrification, in which wealthier residents and upmarket competitors move into a poorer area and displace the venues already there \cite{meltzer2016gentrification, ferm2016preventing}.

Nevertheless, the literature has yet to evaluate, over the longer term, how a mixture of urban features and venue characteristics shapes the evolving popularity trajectories of individual venues. To this end, we ask how far a venue's own popularity trajectory can be predicted by its urban environment. We use a Foursquare dataset covering more than $10^4$ venues in London and Barcelona, each followed for up to 5.5 years. We study these two cities because both have dense, long-established venue scenes and good coverage in the data, yet differ in planning regimes and in their histories of gentrification. We treat each venue's popularity as a time series, segmented into several six-month periods, and compare pairs of series using Dynamic Time Warping (DTW) to account for unequal time series lengths. Grouping venues by this similarity yields six distinct trajectory shapes common to both cities. We then train classifiers to evaluate the feature importance using the permutation importance test. 

Our results show that a neighbourhood's socioeconomic status and density do not predict a venue's trajectory. Instead, a venue's trajectory is mirrored by those of the venues that are in their vicinities, a pattern consistent with a local performance spillover. To measure this directly, we leverage assortativity \cite{Newman2003} to understand whether venues of the same types are more (or less) likely to appear together than chance. Most venues with the same trajectory shape turn out to be neighbours far more often than chance within a few hundred meters. This is a mixing pattern that is only observed on a block- or street-scale, and the pattern weakens as we consider a larger geographical coverage. The effect holds across each city, although the two cities differ in which of the six shapes clusters most strongly. Knowing which shape is located beside which is of use to an owner only if they can tell early which shape their own venue is on, so we further assess how early in a venue's history its trajectory can be told apart from the others, showing that some shapes are recognizable long before the observation period ends while others later. Finally, we offer a qualitative close reading of one rising and one falling district, in London and Barcelona respectively, examining the distinct explanations for spillover under a complex reality of state policies, social histories, and local cultures. Together, these results unveil the complexity behind urban venue performance, and show how neighbourhood vibrancy emerges from particular local conditions and histories. 

\section*{Methods}\label{sec:methods}

We first describe the data and how we turn a venue's raw popularity counts into a trajectory that can be compared across venues. Second, we explain how trajectories are compared with one another and grouped into recurring shapes. Third, we define the four features with which we describe a venue and the area around it. Fourth, we describe the classification and forecasting models, and the relevant evaluation metrics. Fifth, we define a chance-corrected assortativity coefficient, which measures how far a venue's nearest neighbours follow the same shape as it does. Finally, we define our spatial null model, and with how the two case studies were chosen.

\subsection*{Dataset and popularity measure}\label{subsec:data}
\paragraph{Data selection.} Our analyses use a proprietary Foursquare dataset covering two European cities: London, UK and Barcelona, Spain. On Foursquare, users check in to the places they visit and mark the ones they like. We use the likes over time as a primary proxy of popularity for the venues. The total like count of each venue in both cities are heavy-tailed, and thus the values span several orders of magnitude and there is no typical value (see \hyperref[sec:supp_A1]{Supplementary Materials A}). We keep only venues with sufficient data for analysis. Specifically, we select only venues that have existed for at least 130 weeks, corresponding to approximately 2.5 years, and at least 13 likes in total. The process yields $N = 5923$ venues for London and $N = 4718$ for Barcelona. We then evaluated their evolution over a maximum of 5.5 years.  

The clustering and the spatial analyses use all the filtered venues. The classification and forecasting experiments need every feature to be present for every venue, so for robust comparisons we kept only the venues with all features available, resulting in $N = 3694$ and $N = 1139$ in London and Barcelona respectively. Alongside the six-shape label, we also use a coarser, two-way label as a comparative case. A venue is positive when it ended the observation period more popular than it began, that is when $s_i(0) < s_i(T)$, and negative otherwise, to which we refer as the binary task.

\paragraph{Measuring performance.} Following \cite{Yucesoy2016}, we treat success not as something that can be measured in any absolute form but as a ``collective measure, representing a community's reaction to and acceptance of an individual's performance'' (p. 1). Following the urban venue success framework established in \cite{janosov2020}, we aggregate the like counts into 6-month bins for each venue and construct a time series of these counts. The raw number of likes a venue $i$ receives at time $t$ is defined as $\phi_i(t)$. Since raw counts cannot be compared across periods due to the number of users and activity on the platform grows over time, we normalize each raw count by first dividing it by the total popularity across all venues in that same period:

\begin{equation}
 \phi_i^{*}(t) = \frac{\phi_i(t)}{\sum_{j} \phi_j(t)}.
\end{equation}

We then divide by the mean of that quantity, $s_i(t) = \phi_i^{*}(t) / \langle \phi^{*}(t) \rangle$, which allows for a more interpretable metric with a value of $1$ meaning the venue was exactly as popular as the average venue at that moment, and $2$ means twice as popular, and so on. As a normalized metric, $s_i(t)$ will be used the main dependent variable in our analyses. As a comparative case, we also consider the binary success of a venue, where a venue is deemed successful when $s_i(0) < s_i(T)$. 

Because the time series have have different lengths since venues exist during various periods, we align the time series before comparing differences in trajectories. To do so, we use Dynamic Time Warping (DTW) to find the alignment that minimizes the Euclidean distance between any pair of trajectories $s_i(t)$ and $s_j(t)$. Once the minimizing alignment is computed, we apply KMeans clustering to the DTW-aligned time series. Due to the fact that the classifiers and the forecaster described below both take an input of a fixed size, every trajectory is represented as a vector of $T = 11$ six-month periods (i.e., summing up to 5.5 years in total).

\subsection*{Grouping trajectories into recurring shapes}

Venues with the same shape can occur at different speeds and lengths. Therefore, comparing shapes period by period would require alignment between these trajectories. In this work, we use Dynamic Time Warping (DTW) to measure the distance between two popularity shapes after. Specifically, for any pair of trajectories $s_i(t)$ and $s_j(t)$ we use DTW to find the alignment that minimizes the Euclidean distance between them, and we take that minimum as the distance between the pair. We then group venues collectively from London and Barcelona whose shapes are close under this measure using $K$Means clustering, so that every venue in a group follows roughly the same shape. The number of groups is optimized by the gap statistic, a standard rule for choosing how many groups a dataset supports, which identifies $6$ as the optimal number of clusters. Note that the $k$ in $K$Means denotes the number of groups, and is a different quantity from the $k$ used in the classification subsection below for the length of the observed input. The number of groups is fixed at six throughout, and we call it the number of clusters from here on.

\subsection*{Urban features}\label{feature_extraction}

We describe each venue with four features. Two are urban features, including an area's population's socioeconomic status and its population density. Two are venue features, which describe the venue's neighbouring venues, including a venue's uniqueness and the venue type assortativity. 

\paragraph{Deprivation.}
We extracted two urban features attached to the venues in each city. We utilized both data from the original Foursquare datasets, as well as external datasets to compute the features. For London we use the Index of Multiple Deprivation (IMD), a standard measure of how disadvantaged a small area is. A high IMD score means a more deprived area. For Barcelona, we primarily used the average household income from a specific census tracts from the Income Account for Barcelona of the Idescat (Institut d'Estadística de Catalunya) and the Atlas de Distribución de la Renta de las Hogares del INE (Instituto Nacional de Estadística) respectively. As these indices have varying ranges, we rescale the economic status index with the median and inter-quartile range (IQR). As these indices have varying ranges, we rescale the economic status index with the median and inter-quartile range (IQR):

\begin{equation}
    E_i^* = \frac{E_i - E_{\text{median}}}{E_{\text{IQR}}},
\end{equation}

where $E_i$ is the socioeconomic status of the area containing venue $i$. IMD is aggregated using Lower Layer Super Output Areas (LSOAs) and the income level in Barcelona is, on the other hand, aggregated from official census tracts. Both are comparable in terms of population sizes. Because IMD rises with deprivation whereas income rise, we reverse the sign of the Barcelona scores after rescaling, so that in both cities a high value of $E_i^*$ means a less wealthy area. 

\paragraph{Population density.}
We also consider the population density (per hectare) of the areas where the venues are located at, denoted by $D_i$. Due to the different sizes of the census tracts, we again rescale the variables for a more general comparisons between cities. To check for the dispersion of the inequality of $E_i$ and $D_i$, we also calculated the more robust Median Absolute Deviation (MAD). 

\subsection*{Venue features}\label{feature_extraction}

Venues themselves bear inherent characteristics that can catalyze their success or failure. In this light, we compute two metrics that measure these traits beyond the review level. Both are defined over a venue's nearest neighbours, where proximity is defined by a $R-$circle. An $R$-circle of a venue is the disk of radius $R$ meters centered on that venue, and the venues inside it are treated as the venue's neighbours. We use $R = 300$\,m for London and $R = 500$\,m for Barcelona, to account for the different levels of venue density in each city. 

\paragraph{Venue uniqueness.}
First, competition matters to the survival and ratings of venues. With a view to taking into account the uniqueness of the venue within its own area, we first computed the $R-$circle of a venue, which is its circular coverage with a radius of $k$. We then compared the Euclidean distance between the focal venue $i$ and its competitors (i.e. belonging to the same category) within its $R-$circle. Finally, we normalize this count by the average distance between $i$ and all of its neighboring venues. We can the define a venue's uniqueness as follows

\begin{equation}
    u_i = \frac{\langle d(i, j \mid \text{category}_i = \text{category}_j) \rangle_j}{\langle d_{i, j} \rangle_j}.
\end{equation}

A high value means the venue has few same-category competitors close by, and vice versa. 

\paragraph{Assortativity.}\label{sec:assort}
The second venue feature is assortativity, which measures the intensity of venues of the same types co-occuring together spatially. The motivation is that the existence of high performer(s) can also bring positive externalities to its local competitors. It is known that the popularization of specific venues may induce positive and/or negative spillover effects on its neighbouring venues and surrounding areas \cite{duranton2004microfoundations, konishi2003anchor, Daunfeldt2019}. To measure such a spillover effect, we leverage a chance-corrected assortativity coefficient $\hat{\alpha}_i$ for each node $i$, which is given by:

\begin{equation}
 \alpha_i^{(z_i)} = \frac{\displaystyle\sum_{j \in \mathcal{N}_i} 
 \delta(z_j, z_i)}{|\mathcal{N}_i|},
 \label{eq:raw_assortativity}
\end{equation}

where $|\mathcal{N}_i|$ denotes the neighbourhood size, $z$ denotes the group label of a node, which here is the venue's trajectory shape, and $\delta(\cdot)$ is the Kronecker delta function equal to $1$ if node $j$ belongs to group $z_i$ and $0$ otherwise. Concretely, $\alpha_i^{(z_i)}$ is the fraction of venue $i$'s neighbours that share its shape. Spatially, one must define the $R$-circle to obtain the venue's neighbourhood $\mathcal{N}_i$, where $R$ is the radius of the circle of which the venue is the center. We can represent the venues as a co-occurrence network, whereby each venue is a node, and an edge is formed between two venues if and only if one lies within the other's $R$-circle. A venue's set of neighbours $\mathcal{N}_i$ is thus the set of venues it is connected to. Due to a different level of venue density, we applied $R = 300$ for London and $R = 500$ for Barcelona. To account for the expected proportion of connections to any out-group $s$ under random mixing, we define the normalized node-level assortativity index as:

\begin{equation}
 \hat{\alpha}_i^{(z_i)} = \frac{\alpha_i^{(z_i)} - q_{z_i}}{1 - q_{z_i}}, 
 \label{eq:norm_assortativity}
\end{equation}

where $q_{z_i}$ is the global null expected fraction of in-group connection, being

\begin{equation}
 q_{z_i} = \frac{\displaystyle\sum_{j \neq i} \delta(z_j, z_i)}{n - 1}.
\end{equation}

Here $n$ is the number of venues in the city, so $q_{z_i}$ is simply the share of all venues that carry venue $i$'s shape. Similar to general assortative mixing coefficients, $\hat{\alpha}_i^{(z_i)} = 0$ under random mixing, positive when the propensity for in-group connection is higher than expected and vice versa for negative values. The features described earlier use one radius per city. Therefore, positive values indicate assortative behavior (i.e., being connected more often to the same performance group than a different group), and vice versa.

\paragraph{Spatial null model.}

To evaluate the spatial uniformity of the mixing pattern of trajectory types, we devised a spatial null model that places random circles in a given space. Our permutation test starts with randomly choosing $n$ venues and creating a circle of radius $R$ centered on each chosen venue. Within each circle, we take the mean $\hat{\alpha}$ of the venues it contains, and calculate the between-circle variance. A spatially concentrated mixing pattern, in which some areas are far more assortative than others, would produce a high between-circle variance, whereas a spatially uniform pattern would produce a lower one. To determine whether the observed variance is greater than would be expected by chance, we compare it against a null distribution generated by reshuffling the trajectory type labels across venues $N = 1000$ times while holding their coordinates fixed, recomputing the between-circle variance for each shuffle. The one-sided $p$-value is the proportion of shuffled variances that is equal to or exceed the observed variance, that is:

\begin{equation}
    p = \frac{1 + \sum_{s=1}^{N} \mathbf{1}\left[\sigma^2_{\text{null},s} \geq \sigma^2_{\text{obs}}\right]}{N + 1}.
\end{equation}

To avoid unstable estimates in sparsely populated circles, we discard any circle containing fewer than five venues, which leaves between $327$ and $471$ usable circles depending on the city and the radius used. In the main analysis, we set $n = 500$ m for London and $R = 300$ m for Barcelona. We also repeat the test at $R = 200$, $250$, and $500$ m as a robustness check (see \hyperref[tab:random_circle_test]{Supplementary Materials C.5 Table 10}). We treat $p < 0.05$ as significant, and note that the test is run at four radii in each of two cities without correction for the number of tests.

Finally, to evaluate the pairwise co-occurrence of shapes, we construct a mixing matrix $M$, where each cell $M_{ab}$ is the observed co-occurrence between any given pair trajectory shapes $a, b$. We compute the ratio of $M_{ab}$ to the expected co-occurrence under random mixing $E_{ab}$, which is given by 

\begin{equation}
E_{ab} =
\begin{cases}
2 \, p_a p_b \, L, & a \neq b, \\
p_a^2 \, L, & a = b,
\end{cases}
\end{equation}
where $p_a$ and $p_b$ denote the fraction of venues of shape $a$ and $b$ respectively, and $L$ denotes the total number of neighbouring venue pairs. For interpretability purposes, we take the $\log_2$ of the ratio

\begin{equation}
S_{ab} = \log_2\left(\frac{M_{ab}}{E_{ab}}\right),
\end{equation}

such that $S_{ab} = 0$ means exactly as often as chance, $S_{ab} > 0$ means more often than chance, and $S_{ab} < 0$ means less often.

\paragraph*{Review Feature.}
Finally, we extract the reviews of all venues across its lifespan and focus on the reviews' sentiment. We first annotated the sentiment polarity using a DistilBERT-based multilingual sentiment analysis model due to the existence of non-English comments in our datasets, particularly in the Barcelona sample. We then recoded the sentiment label into a 5-point Likert scale $\gamma \in \{-2, -1, 0, 1, 2\}$, from strongly negative to strongly positive.  We computed the mean $\gamma$ of each venue on a six-monthly basis and constructed a vector of a normalized length $T = 11$ using linear interpolation and extrapolation for venues with different ages. Missing data were imputed using the mean value of the vector. We report the descriptive statistics of the sentiment values in \hyperref[fig:supp_sentiment_time]{Supplementary Materials Figure S5}. Note that the sentiment feature is only used for the early classification task given that sentiment feature is a time series vector and is secondary to our main research questions regarding urban surroundings of venues, we therefore do not report regarding the sentiment in the main results. 

\subsection*{Classification and Forecast Models}

\begin{figure}[htp!]
 \centering
 \includegraphics[width=\linewidth]{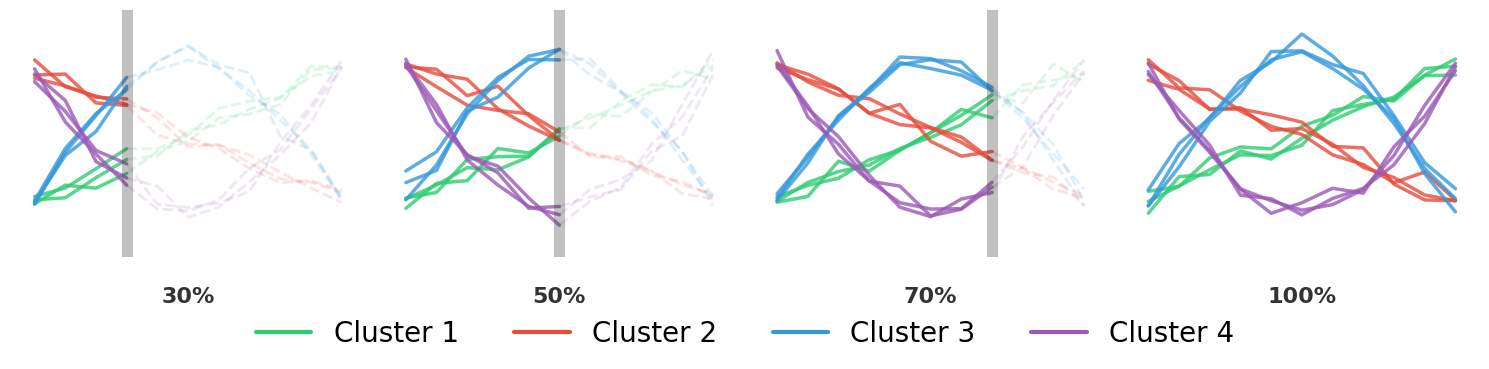}
 \caption{Toy example of early classification of venue popularity trajectories. We test at which point of time $t$ (denoted as the percentage of data available to the model in the figure) the cost is minimized. Intuitively, for instance, trends are nearly indistinguishable when the data completeness is insufficient and if they are moving into the same direction (cluster 1 and cluster 3; cluster 2 and cluster 4 at 30\%). It is only when more data are revealed (at 70\%), we are able to distinguish the actual shapes of these paths. Notably, when the completeness is 100\%, the task is the same as a detection task.}
 \label{fig:toy_example}
\end{figure}

\paragraph*{Classification model.} The goal of our classification task is to understand how early we can reliably classify the trajectory during an early onset for a better decision-making process. Using the features discussed above, we can formulate the problem as a sequential decision-making task, wherein early classification is made in order to predict which trajectory shape the given time series data is going to match, a problem domain generally defined as Early Time Series Classification (ETSC). While there is a host of ETSC algorithms (for details, see the comprehensive reviews in \cite{Gupta2020, renault2025early}), we do not directly apply any ETSC algorithm. Instead, we conducted a series of exploratory analyses to understand the trade-off between earliness and accuracy with four classical machine learning models, including Multinomial Logistic Regression, Random Forest, and XGBoost. We conducted a stratified 5-fold 80-20 split on the original dataset, and trained using varying levels of data incompleteness. The parameter $k \leq T$ controls the number of data points that the model is able to see. If $k=T$, then the model has complete access to the data, otherwise the model is trained on only $\frac{k}{T}$ of the full data, which is equivalent to the truncated series $s_i(0,k)$. Note that each $k$ period represents a six-month period.

We run the classifiers in two settings. In the trajectory-only setting, each classifier is shown only $s_i(0,k)$, and must classify the input as one of the trajectory shapes. In the full-feature setting, we add the urban and venue features to the same models and rerun them. We then tested different combinations of features to assess how much each contributes at varying time points $k$.

\paragraph*{Forecasting model.}
For the forecasting model, we predict the actual normalized popularity values of a given trajectory. Here, we trained a Seq2Seq model using a Gated Recurrent Unit (GRU)-based Encoder-Decoder architecture for exact value forecasting. Given the size of the time series data is small, the encoder-decoder architecture is relative simple to avoid overfitting. A bidirectional GRU encoder was used to capture temporal dependencies in both forward and backward directions, followed by a fully connected decoder that maps the learned representation to predicted values. By experimenting with more complex architecture, we found that additional layers resulted in overfitting with negative $R^2$ in the test set, and thus confirming our choice of a simpler setup. Two training strategies were implemented. A joint training where a single model is fitted to all venues, whereas the per-cluster training fits a separate forecaster to each trajectory cluster using the same architecture.

\subsection*{Qualitative case study}

According to the principle of Most Similar Systems Design \cite{Anckar2008}, we select two cases (i.e., Hackney in London and Ciutat Vella in Barcelona) because they are similar on the dimensions our quantitative models found unpredictive given that both rank among the most deprived, densely populated districts in their respective cities (see \cite{HackneyCouncil2025, SnchezAguilera2021}) and experience decades of policy-driven renewal and gentrification (see \cite{ferm2016preventing, PascualMolinas2009}), yet their venue clusters diverged sharply in trajectories. Crucially, we remind the reader that here we do not attempt to establish direct statistical causality between longitudinal social metrics and venue performance. Instead, we utilize district histories and social theories to interpret the underlying mechanisms behind these micro-spatial spillovers, which serve as exemplary cases of street-level dynamics. As urban geographer Flyvbjerg demonstrates, detailed case study of exemplars is essential for theorizing processes that large-scale, aggregate metrics may neglect \cite{Flyvbjerg2006}.

\section*{Results}

\subsection*{Venue popularity follows six recurring shapes across cities}

\begin{figure}[ht]
 \centering
 \includegraphics[width=\linewidth]{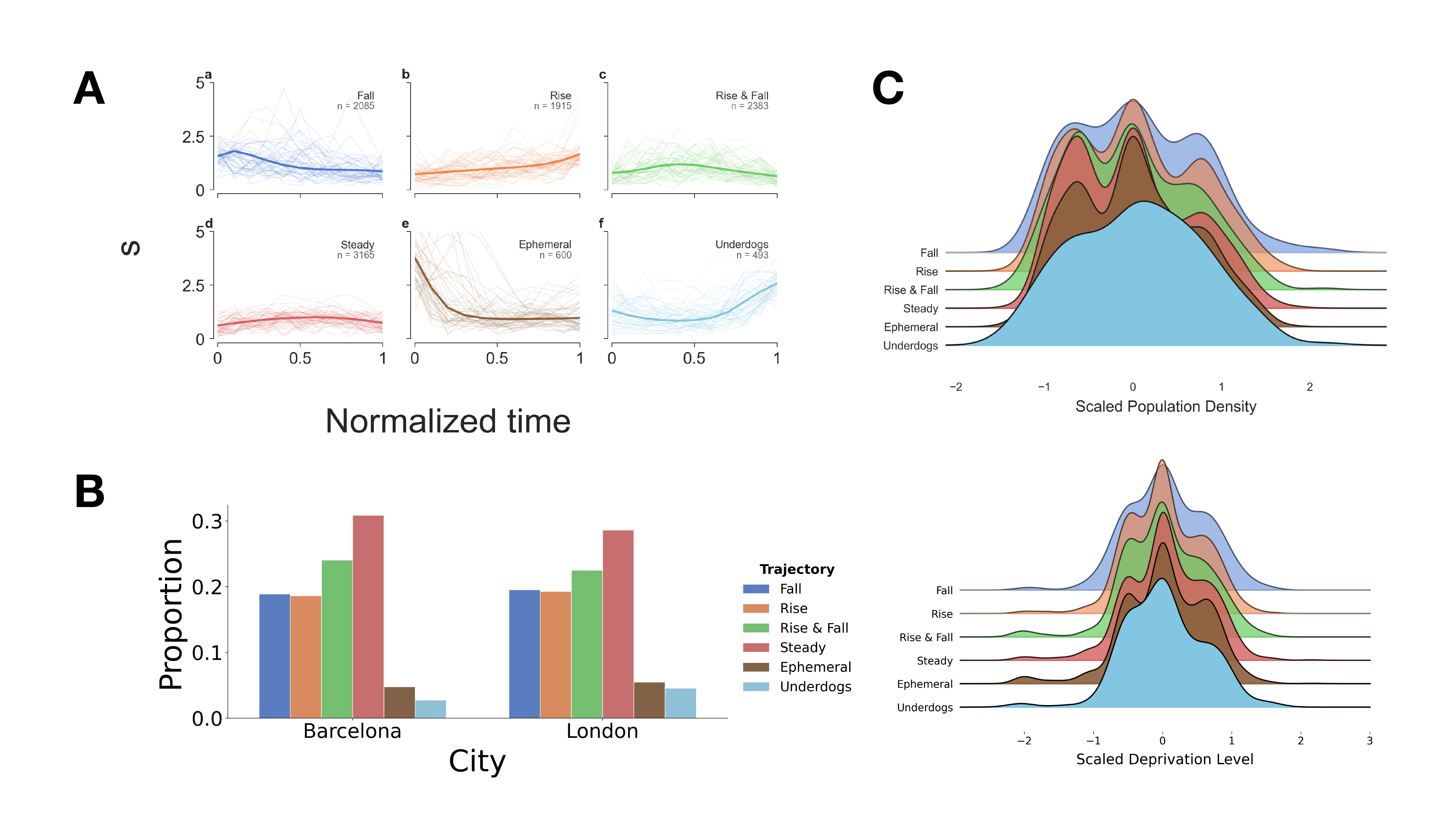}
 \caption{Typology of popularity trajectories of urban venues in London and Barcelona. \textbf{Panel A}: the general shapes of the six recurring trajectory types. \textbf{Panel B}: proportional count of each type in Barcelona and London. Both cities have a similar distribution, with Steady being the most common type and Underdogs the least common. \textbf{Panel C}: the empirical distribution of the two urban features, namely population density and scaled deprivation level, grouped by trajectory types.}
 \label{fig:fig1}
\end{figure}

Grouping venues by the shape of their popularity shape (see \hyperref[sec:methods]{Methods}) recovers six dominant shapes: (a) Fall, (b) Rise, (c) Rise \& Fall, (d) Steady, (e) Ephemeral, and (f) Underdogs (see \hyperref[sec:supp_clusters]{Supplementary Materials B} for details). The same six shapes appear in both London and Barcelona. These clusters exhibit very clear patterns in terms of their evolution (see \hyperref[fig:fig1]{Figure~\ref{fig:fig1}}), and split into near-linear and non-linear trajectories.

The near-linear shapes, i.e. the Fall, Rise and Steady clusters, are generally expected patterns. The non-linear ones capture more complex shifts in venue popularity. A venue in the Ephemeral cluster was popular in its first years of trading but lost that popularity quickly. Underdogs are the mirror image: they start out unnoticed and gain a substantial following by the end of the observation period. Rise \& Fall venues trace a mild inverted U, growing in popularity up to roughly the midpoint of their lifespan and then declining over the second half. Cluster sizes range from 493 venues (Underdogs) to 3,165 venues (Steady).

Notably, no U-shaped Fall \& Rise cluster emerged in either city. Recovery after a sustained decline is therefore not one of the recurring shapes in our data. The inverse shape, Rise \& Fall, is present and populous, so the asymmetry is not an artifact of what the clustering can represent: in our data, venues that lose popularity do not form a recurring pattern of coming back. Recovery, when it happens, is either rare enough or idiosyncratic enough not to constitute a shape of its own. Steady and Rise \& Fall are the least visually separable pair, so we compared their summary statistics directly. The Steady cluster has a lower standard deviation and a lower average curvature, meaning that its shapes bend less and run straighter, than Rise \& Fall, confirming their differences (\hyperref[fig:supp_traj_shape_std]{Supplementary Materials B, Figure S5})

With the six shapes in hand, we test whether a venue's neighbourhood can predict which shape a venue belongs to. We answer this with the classification experiments described in \hyperref[sec:methods]{Methods}. Urban features alone are not predictive, in either the binary or the six-class setting (\hyperref[tab:binary_results]{Supplementary Materials C.2, Table 5} and \hyperref[tab:multiclass_results]{Supplementary Materials C.2, Table 6}). Venue features do carry signal, reaching F1 scores of $0.81$ on the binary task and $0.71$ on the six-class one, whereas models given only the urban features reach $0.10$--$0.21$ on the same six-class task. We note that, nonetheless, the relationship between venue features and trend type is likely non-linear as logistic regression performs below baseline while tree-based models managed to make accurate predictions. Moreover, we found in our sample that London venues are easier to predict than Barcelona ones in both the binary and the six-class case \hyperref[tab:city_comparison]{Supplementary Materials C.2, Table 7}).

Across all non-linear models and both cities, cluster assortativity (see definition in \hyperref[sec:methods]{Methods}) is the strongest predictor of a venue's trajectory type (see \hyperref[tab:feature_import]{Supplementary Materials C.3 Table 8}). In turn, our results demonstrate that the tendency to which a venue is clustered with, or not, other venues belonging in the same trajectory type is decisive for classifying the venue's own type. In line with longstanding work on agglomeration economies \cite{glaeser2010agglomeration}, such a finding suggests that sustained popularity patterns exhibit some kind of spatial spillover effect, whereby geographically clustered venues rise or decline together, which we will further explore in the upcoming sections. Yet the weak predictive power of population density and socioeconomic status suggests the link between neighbourhood characteristics and venue success is less direct than conventional theories of neighbourhood effects assume \cite{durlauf2004neighbourhood, vanham2012neighbourhood}. These patterns, however, are established post-hoc, once a venue's full trajectory is known. This leaves open a practical question for business owners and planners: how early in a venue's life can its trajectory be classified? We address this in the next section.

\subsection*{How early a venue's performance is visible depends on when its signature appears, not how complex the shape is}
\paragraph{Early classification of trajectory type}
In reality, business owners must infer what they should and should not do from partial information about their past and current performance, and do so as early as possible. We begin by establishing an upper bound of predictability, that is how well can these models do when the full time series is visible (i.e., $k = T$)? As expected, labeling a venue once its entire history is visible is easier than predicting from a partial one, with most models exceeding $0.70$ accuracy. Note that even at $k = T$ the task is not trivial, since the model sees only the venue's own shape and must recover from it a grouping that was formed by comparing whole shapes against one another. The best performing model is XGBoost, reaching a macro F1 of $0.83$, far above the majority-class and random baselines given in \hyperref[sec:methods]{Methods}. It is consistently the best performer across all clusters (\hyperref[tab:perclass_comparison]{Supplementary Materials C.1, Table 4}). Examining the performance per class, our results revealed that trajectories in the Ephemeral cluster are the easiest to classify (mean $F1 = 0.86$) across all models, meanwhile Rise \& Fall is the most difficult to classify (mean $F1 = 0.54$), followed by Underdogs (mean $F1 = 0.56$).

\begin{figure}[htp!]
 \centering
 \includegraphics[width=\linewidth]{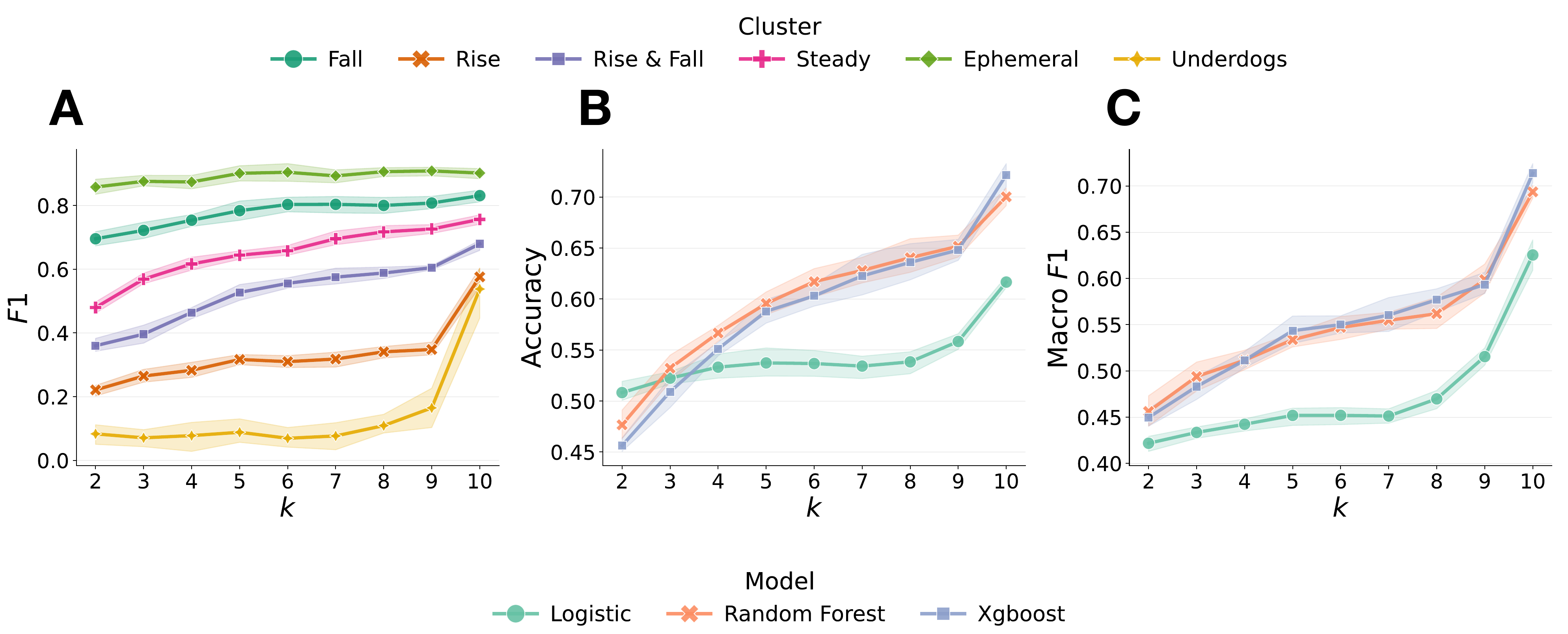}
 \caption{Earliness-accuracy trade-off of early classification. \textbf{Panel (A)} shows the main result of XGBoost, the best performing model overall. Categories such as Fall and Steady do not gain much over $k$, meaning that most decisions can be made earlier given their values. However, other categories that are more difficult classify in the first place. \textbf{Panel (B) - (D)} are the learning curves with increasing $k$ of different models.}
 \label{fig:fig2}
\end{figure}

As we have established the upper bound, we now report what happens as less data is available, a problem known as the earliness-accuracy trade-off of trajectory classification with partial data. \hyperref[fig:fig2]{Figure 3} demonstrates the performance of the three models with varying values of the input richness parameter $k$. As we expected, how much accuracy is given up by deciding early differs sharply between shapes. For instance, the Fall trajectory has a relatively stable F1-score across different values of $k$, showing only gradual improvement with the increase of $k$, as its declining trend is already apparent early in the sequence. The cost of deciding early is therefore not the same for each venue because of different predictability of the shapes. In particular, Rise and Rise \& Fall are, by definition, indistinguishable in earlier time points, as both begin with an increasing trend, which is reflected in the F1-score of Rise remaining flat until $k=9$ before increasing sharply at $k=10$. A similar dynamic applies to Underdogs, whose defining sharp increase in popularity occurs at the very end of the sequence, and thus its F1-score only starts increasing when $k\geq8$. All models show a similar rising trend in performance when averaged over all clusters (\hyperref[fig:fig2]{Figure 3B and C}), although XGBoost was able to improve earlier than Random Forest and Logistic Regression and had a sharper increase in performance despite having similar initial performance.

Thus far, the models have only observed venue's own shape. We then asked whether adding the urban and venue features defined in \hyperref[sec:methods]{Methods} improves the predictability of these trajectories. Consistent with our earlier finding that urban context does not predict trend type, the added features did not help. Adding urban and venue-specific features had, at best, a marginal and inconsistent effect, whether performance is averaged over all venues or measured separately within each trajectory type. For instance, some classes improve slightly while others deteriorated (\hyperref[fig:early_class_ablation_by_k]{Supplementary Materials C.4 Figure S7}). The one case worth noting is Underdogs, one of the two hardest types to classify, for which combining all available features offered a small improvement over the shape-only models in the earliest, most information-poor prediction windows ($k=2$-$3$), though the difference remains meager. The picture to carry forward is therefore that what a venue is and where it sits add little to what its own shape so far already says.

\paragraph{Forecasting of venue popularity values} Classifying the shape early, however, is only part of the problem. A venue owner needs to know not just which path they are on but also its duration and intensity. In the remainder of this section, we turn to forecasting the actual popularity values. Training a separate forecaster for each trajectory shape (P in \hyperref[tab:prediction_table]{Table 1}, worked slightly better than training a single forecaster for all venues (J in \hyperref[tab:prediction_table]{Table 1}), although the gap was modest (mean MSE $0.171$ vs. $0.190$). The advantage of per-cluster training was concentrated at smaller $k$, i.e. at shorter input lengths and among the harder clusters.  

Regarding the earliness-accuracy trade-off, we again found that the forecast error increases with the decrease in data input length, i.e. less the data the model sees, the harder its future is to predict. As the prediction lead time decreases (from 30 months, $k = 5$ to 6 months, $k = 10$), the overall error falls from $0.224$ to $0.099$. Under joint training, error remains flat at approximately $0.20$ for most lead times before dropping to $0.121$ when predictions are made 6 months ahead. Predictability also varied sharply by shape. Steady venues were the most forecastable at every input length but the longest, while Underdogs were the least forecastable overall, with error several times that of any other cluster under joint training. Notably, per-cluster training narrows this gap substantially at $k = 9$ and $k = 10$. Underdogs depend on a late reversal that the model cannot anticipate, so the capacity to turn a declining venue around is largely unpredictable from past popularity alone, resting on factors absent from our data, such as chance, wider market shifts and changes in management.

All in all, the prediction results reflect an uncomfortable reality of business decision-making. As the forecast horizon lengthens, their performance becomes more unpredictable. In other word, the earlier an answer is sought, the less trustworthy that answer is. The more constructive implication is that forecasts at a six-month lead remain reasonably accurate (see \hyperref[prediction_table]{Table 1}, $k = 10$), giving decision-makers a short but practical window to strategize and plan. 

\begin{table*}[h]
\centering
\begin{tabular}{cl|cccccc|c}
\toprule
\textbf{$k$} & & \textbf{Fall} & \textbf{Rise} & \textbf{Rise \& Fall} & \textbf{Steady} & \textbf{Ephemeral} & \textbf{Underdogs} & \textbf{Overall} \\
\midrule
\multirow{2}{*}{5} & J & 0.234 & 0.268 & 0.199 & \cellcolor{best}0.103 & 0.207 & \cellcolor{worst}0.848 & 0.212 \\
& P & 0.225 & 0.211 & 0.165 & \cellcolor{best}0.087 & 0.207 & \cellcolor{worst}0.450 & 0.224 \\
\cmidrule{2-9}
\multirow{2}{*}{6} & J & 0.191 & 0.277 & 0.182 & \cellcolor{best}0.096 & 0.189 & \cellcolor{worst}0.845 & 0.200 \\
& P & 0.181 & 0.199 & 0.140 & \cellcolor{best}0.080 & 0.193 & \cellcolor{worst}0.384 & 0.196 \\
\cmidrule{2-9}
\multirow{2}{*}{7} & J & 0.191 & 0.274 & 0.173 & \cellcolor{best}0.107 & 0.225 & \cellcolor{worst}0.728 & 0.197 \\
& P & 0.173 & 0.199 & 0.116 & \cellcolor{best}0.084 & 0.224 & \cellcolor{worst}0.373 & 0.195 \\
\cmidrule{2-9}
\multirow{2}{*}{8} & J & 0.156 & 0.291 & 0.174 & \cellcolor{best}0.121 & 0.189 & \cellcolor{worst}0.720 & 0.198 \\
& P & 0.132 & 0.201 & 0.106 & \cellcolor{best}0.075 & 0.195 & \cellcolor{worst}0.261 & 0.162 \\
\cmidrule{2-9}
\multirow{2}{*}{9} & J & 0.199 & 0.318 & 0.183 & \cellcolor{best}0.115 & 0.226 & \cellcolor{worst}0.717 & 0.211 \\
& P & 0.109 & 0.192 & 0.071 & \cellcolor{best}0.062 & \cellcolor{worst}0.242 & 0.222 & 0.150 \\
\cmidrule{2-9}
\multirow{2}{*}{10} & J & \cellcolor{best}0.062 & 0.192 & 0.123 & 0.092 & 0.105 & \cellcolor{worst}0.305 & 0.121 \\
& P & 0.059 & 0.088 & \cellcolor{best}0.050 & 0.062 & \cellcolor{worst}0.202 & 0.132 & 0.099 \\
\midrule
\multirow{2}{*}{\textbf{Mean}} & J & 0.172 & 0.270 & 0.172 & 0.106 & 0.190 & 0.694 & 0.190 \\
& P & 0.147 & 0.182 & 0.108 & 0.075 & 0.211 & 0.304 & 0.171 \\
\bottomrule
\end{tabular}
\caption{Forecasting performance, measured as mean squared error (MSE). $k$ indexes historical observations in decreasing recency, from $k=10$ (6 months prior) to  $k=5$ (5 years prior), so a smaller $k$ means an earlier and less informed forecast. Given the input length $k$, the model predicts the remaining values of a performance trajectory after time $k$ the exact values from $k+1$ until $k = T$. J = joint training, a single model trained on all venues; P = per-cluster training, a separate model trained for each trajectory shape. The six named columns shows the MSE of each trajectory type, and \textbf{Overall} gives the mean of all venues for a particular time $k$. Within each row, the lowest error is shaded \colorbox{best}{green} and the highest \colorbox{worst}{red}. As hypothesized, the difficulty of trend forecasting grows as $k$ decreases, meaning that long-term prediction is more challenging than short-term prediction. We also found that Underdogs are the most difficult to predict, meanwhile the Steady cluster is the easiest to predict across even all $k$.}
\label{tab:prediction_table}
\end{table*}

\subsection*{Trajectory types mix at the block scale, not the neighbourhood}\label{sec:mixing}

The preceding classification experiments identified a venue's spatial assortativity as the strongest predictor of its trajectory type. However, we have yet to establish, empirically, whether venues of the same type actually mix together in space, at what scale this mixing occurs, or whether the pattern reflects a universal tendency for (anti-)homophily across the city. To do so, we measure the chance-corrected assortativity metric $\hat{\alpha}_i$ (see \hyperref[sec:assort]{Methods}) such that $\hat{\alpha}_i = 0$ corresponds to a neighbourhood no more concentrated than chance, $\hat{\alpha}_i > 0$ to an excess of same-type neighbours, and $\hat{\alpha}_i < 0$ to the opposite.

\begin{figure}[htbp!]
 \centering
 \includegraphics[width=\linewidth]{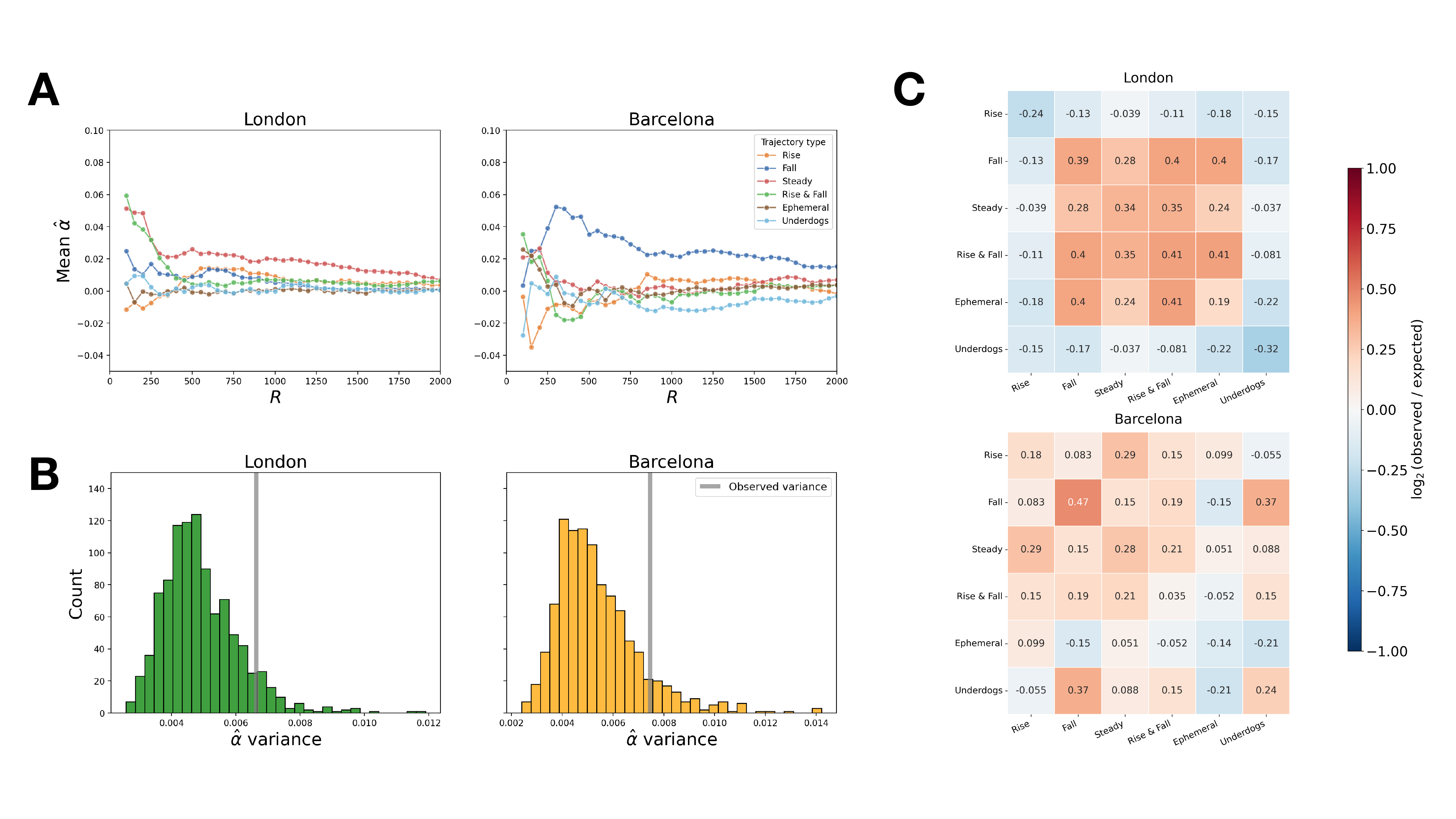}
 \caption{\textbf{Panel A:} The mean chance-corrected assortativity coefficient $\hat{\alpha}$ is scale-variant in both London and Barcelona, with most venues experiencing stronger (dis)assortativity on the block-level, but decays toward zero due to the inclusion of more venues as $R$ increases. \textbf{Panel B:} Results of the spatial shuffle null model. The histograms show the distribution of the between-circle variance $\sigma_{\text{null}}^2$ generated from the null model with $R = 300$, and the gray vertical line is the observed between-circle $\sigma_{\text{obs}}^2$. \textbf{Panel C:} For each pair of trajectory types, the heatmap shows how often venues of those two types are found within $R = 300$\,m of each other. This count is divided by the count expected under a random reassignment of labels to the same venue locations, and shown on a $\log_2$ scale. Values above zero mean the pair co-occurs more than chance would give, values below zero less. Note that this panel uses $R = 300$\,m for both cities, unlike the per-venue assortativity feature, which uses $300$\,m in London and $500$\,m in Barcelona.}
 \label{fig:mixing}
\end{figure}

First, we found that the mixing pattern is scale-variant. (Dis)assortativity is a block-level phenomenon that vanishes with scale, and the exact pattern is trajectory type-specific. In both cities the signal is most distinguishable and stable at radii of roughly $200$--$600$ m (mean $\hat \alpha \approx 0.008$ for both cities at $300$ m) and decays toward zero beyond $1$ km, as $\hat{\alpha}$ will, by construction, converge to the city-level mean as $R$ goes up. Nevertheless, it was found that the trajectory types have distinct mixing patterns (see \hyperref[fig:mixing]{Figure 4A}). While Steady is the most spatially homophilic in London, it is the Fall venues that are the most likely to be clustered together with their own types in Barcelona. As such, the clustering effect of venue popularity appears to be scale-, type-, and context-dependent.

Next, we tested whether this mixing is uniform across space within a city using the random-circle permutation test described in \hyperref[sec:methods]{Methods}. In London, the observed variance is indistinguishable from the label-shuffled null across all radii tested (see \hyperref[fig:mixing]{Figure 4B}), indicating that the mixing tendency does not vary meaningfully between neighbourhoods but is a general feature of the city. In Barcelona, the observed variance consistently exceeds the null by a wider margin than in London, however this is only significant when $R = 200$ ($p = .022$). Since we ran this test at four radii in each of the two cities without correcting for the number of tests, one significant result out of eight is close to what we would expect by chance alone, and we therefore do not conclude it as evidence that Barcelona's mixing is concentrated in particular areas (see detailed results in \hyperref[tab:random_circle_test]{Supplementary Materials C.5}). Overall, we found weak to no evidence that the mixing pattern is spatially concentrated, and instead it is a generalizable phenomenon across neighbourhoods in both cities. 

Finally, given the mixing patterns, we examined which types each shape tend to co-occur together using the observed-to-expected adjacency ratios at $R = 300$ m (see \hyperref[fig:mixing]{Figure 4C}). Two patterns recur across both cities. Most values on the diagonal are above zero on $\log_2$ scale, i.e. above a ratio of $1$, confirming that venues of most types are found near their own kind more often than chance would give at the block scale. However, the exceptions are Rise and Underdogs in London and Ephemeral in Barcelona, whose diagonal entries fall below zero. More tellingly, the Underdogs and Ephemeral venues behave differently from the rest. In London, Underdogs are underrepresented as neighbours of nearly every other type, including themselves. In Barcelona the pattern is different, that is Underdogs are strongly over-represented both among themselves and alongside Fall venues. On the other hand, Ephemeral venues are under-represented as neighbours of nearly every other type in Barcelona, whereas in London they are more likely to be clustered with Fall and Rise \& Fall, alongside other Ephemeral venues. The remaining types, Fall, Steady and Rise \& Fall, are not only over-represented with themselves but also strongly co-occur with one another in London, suggesting that these three trajectories share block-scale neighbourhoods rather than segregating by type.

All in all, our results suggest that the neighbourhood effect is not organized entirely in the same way in the two cities. The type that most strongly clusters with its own kind differs, being Rise \& Fall in London and Fall in Barcelona, and Barcelona shows a consistently, if weakly, more concentrated mixing field than London. Yet the qualitative structure is shared. Mixing is a block-level, type-specific phenomenon that is broadly general across each city rather than confined to particular districts. Overall, these findings point to a spillover-like process operating at the scale of the street rather than the whole neighbourhood, whose strength and expression nonetheless depend on the particular urban context in which it is found. 

\subsection*{A comparative case study of venue clusters in Hackney and Ciutat Vella}

\begin{figure}[htp!]
 \centering
 \includegraphics[width=\linewidth]{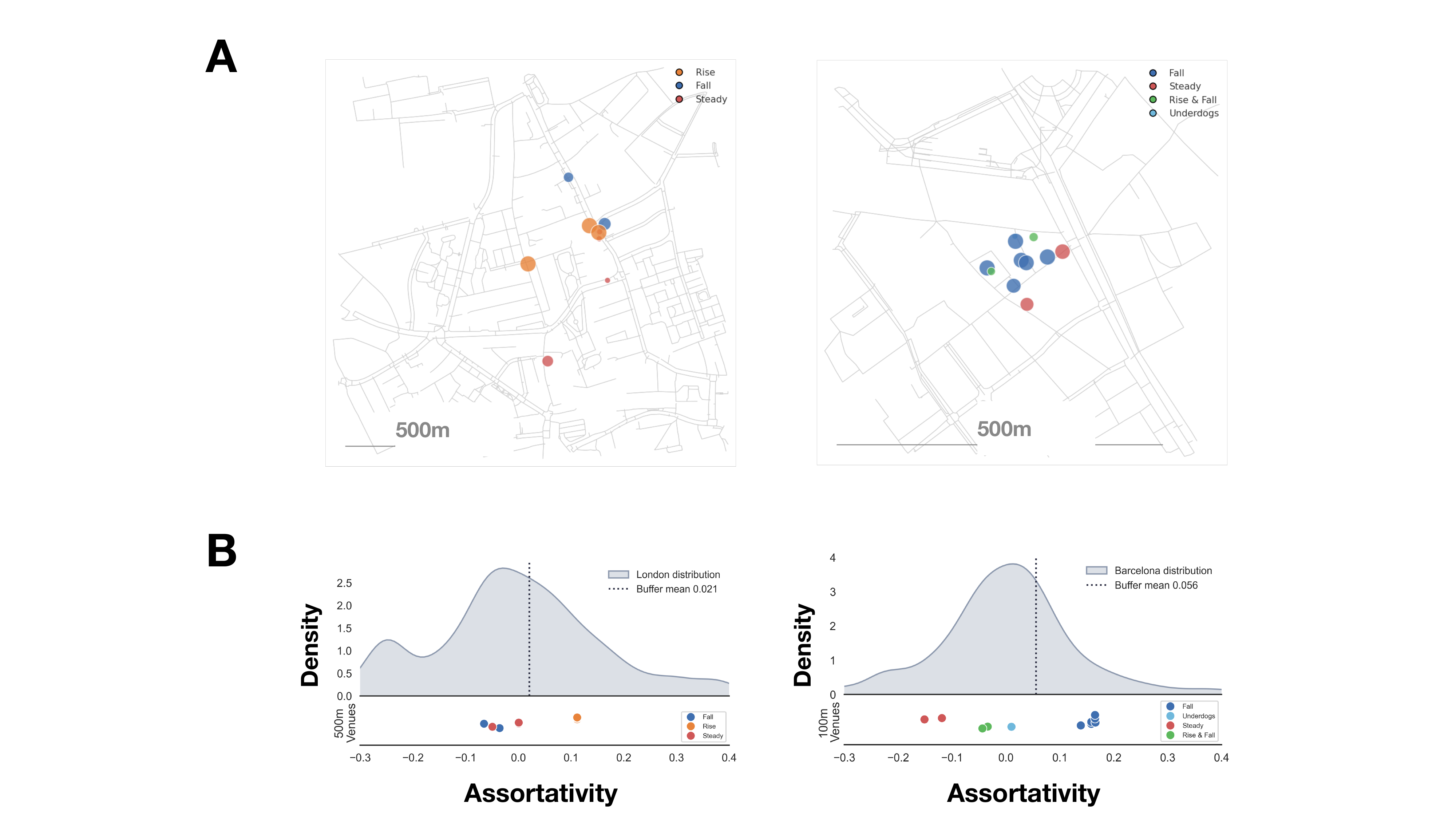}
 \caption{\textbf{Panel A:} Street map of the focal areas investigated in London (left) and Barcelona (right). The size of the points indicates the assortativity (scaled for clarity) and the color indicates the trajectory types. \textbf{Panel B:} Upper panel shows the distribution of assortativity in both cities with the KDE plot (gray) showing the city average and the dotted vertical line indicates the mean assortativity of the focal zones. Lower panel shows the exact values of assortativity of each venue in the focal areas.}
 \label{fig4}
\end{figure}

Why should two venues a few hundred meters apart share similar performance? One candidate explanation is a spatial spillover effect, whereby nearby venues gain or lose popularity together through their exposure to shared local social and economic conditions \cite{Yoshimura2020}, and our quantitative results are consistent with such an effect operating largely at the street level. To go further, this section explores qualitatively why aggregate administrative units (e.g., census tracts) cannot capture micro-spatial dynamics accurately. The two cases are a Rise cluster in the London Borough of Hackney, and a Fall cluster in the Ciutat Vella district of Barcelona.

\paragraph{Boutiquing and residential gentrification of hackney}
In Hackney, we find tight spatial clusters of Rise venues, confined to a small, highly localized area, whose average assortativity ($\hat{\alpha} = 0.112$ -- $0.150$, see {\hyperref[tab:comparative_venues]{Table 2}) is higher than both London's mean and median . Hackney's broader trajectory provides the historical context for why these street-level spillovers may emerge. Among the fastest-growing economies in London \cite{HackneyCouncil2025}, its trajectory is rooted in early 2000s affordable workspace policies designed to retain creative professionals, alongside post-2012 Olympic regeneration and Overground transit expansions \cite{ferm2016preventing, Lagadic2019}. This repositioning attracted young professionals, establishing the borough as a `middle-class hipster' residential-commercial hub \cite{Butcher2016} where shifting neighbourhood taste dislocated traditional businesses through commercial rent inflation.

At the micro-scale, this dynamic is visible along Clarence Road and Lower Clapton Road. Take the $300$\,m circle that Methods defines around one venue there, a cooperatively run vegan cafe that is a part of the local punk scene. Almost every Rise venue in our Hackney sample falls inside that single circle, and they sit within a few doors of one another along the same two streets, including the cafe itself, an independent wine bar, and a branch of a premium hair-salon chain. Table \ref{tab:comparative_venues} lists these venues together with every other venue in the same circle.These venues exemplify Deborah Cowen's concept of `hipster urbanism' \cite{cowen2006hipster} and Hubbard's commercial `boutiquing' \cite{hubbard2017battle}, transforming working-class aesthetics into curated `shabby chic' consumption spaces \cite{Halnon2002}. 

Because boutiquing relies on contiguous window-shopping and shared foot traffic, positive spillovers remain hyper-localized rather than spreading across the broader neighborhood. In our sample, these growing venues co-locate with established venues born out of gentrification (e.g., chain hipster pizza shop, craft beer retailers), while standing in contrast to the adjacent heritage and community venue that belongs to the Fall shape. Despite extensive gentrification, Hackney remains segregated \cite{stonehirsch2019local, mhclg2020indices}, combining severe child poverty with London's third most expensive housing market \cite{trustforlondon2026}. Severe deprivation and booming commercial vibrancy thus can coexist within the same neighborhood, and is only disentangled cleanly when evaluated at the micro-street level.

\paragraph{Tourism-driven gentrification in Ciutat Vella}
On the contrary, the Ciutat Vella / Upper Raval area in Barcelona reveals a contrasting trajectory driven by tourism gentrification and mismanaged urban policy. Hosting over 10 holiday homes per hectare, nearly five times the city average, Ciutat Vella experienced massive rental price increase over the years and 605 registered evictions in 2017 alone \cite{SnchezAguilera2021}. This mirrors Hackney's commercial transformations, but targets non-residential consumers while vastly neglecting residential needs \cite{JimnezPacheco2023}. Local residents remain segregated by urban policies rooted in late-1970s revitalization models \cite{lopezsanchez1986} and post-2008 economic austerity.

The two districts diverge significantly in timing and history. Ciutat Vella's transformation began ahead of the 1992 Olympics \cite{PascualMolinas2009}, coinciding later with post-2008 economic stagnation to incubate speculative tourism gentrification rather than resident-led commercial growth \cite{CocolaGant2018, Domnguez2017}. This matches what Aalbers calls fifth-wave gentrification \cite{Aalbers2018} --- neighbourhood change driven less by middle-class residents moving in and more by investors treating rental housing as a financial asset. The proliferation of short-term lets, via global platforms such as Airbnb, directly undermines the regular local consumer base for independent storefronts \cite{Sans2016, Lagonigro2020}.

Empirically, this dynamic is also observable at the street level. Within a compact retail area in Upper Raval, we identify a spatial cluster in which a majority of venues are classified as Fall with assortativity scores ($\hat{\alpha} = 0.139$--$0.166$) higher than the city's average (see \hyperref[fig4]{Figure 4B}). Long-standing local stores are disproportionately represented among these venues. For instance, a longstanding bakery ($\hat{\alpha} = 0.139$) and a record shop ($\hat{\alpha} = 0.165$), both classified Fall, exemplify a broader pattern in which culturally embedded, resident-oriented businesses deteriorate collectively rather than in isolation. Similar to boutiquing, tourist foot traffic concentrates also along specific areas and shops rather than diffusing evenly through the urban fabric. Commercial displacement therefore is not district-wide but as a street-localized phenomenon.

\paragraph{The complexity of micro-spatial spillover effects}

\begin{table}[htbp]
\centering
\small
\setlength{\tabcolsep}{5pt}
\begin{tabularx}{\textwidth}{>{\centering\arraybackslash}p{2.2cm} >{\centering\arraybackslash}X >{\centering\arraybackslash}c >{\centering\arraybackslash}c >{\centering\arraybackslash}c}
\toprule
\textbf{City} & \textbf{Venue Category} & \textbf{Trajectory} & \textbf{Status} & \textbf{Chain?} \\
\midrule
\multirow{9}{*}{\textbf{London}} 
 & Premium hair salon & \multirow{3}{*}{Rise} & \multirow{8}{*}{Open} & \textbf{\checkmark} \\
 & Wine bar & & & \textbf{\textsf{X}} \\
 & Vegan cafe & & & \textbf{\textsf{X}} \\
\cline{2-3} \cline{5-5}
 & French brasserie \& cocktail bar & \multirow{4}{*}{Steady} & & \textbf{\textsf{X}} \\
 & Craft beer retailer & & & \textbf{\checkmark} \\
 & Hipster pizza & & & \textbf{\checkmark} \\
 & Ramen store & & & \textbf{\checkmark} \\
\cline{2-3} \cline{5-5}
 & Historic event center & \multirow{2}{*}{Fall} & & \textbf{\textsf{X}} \\
\cline{4-5}
 & Record store & & Permanently Closed & \textbf{\textsf{X}} \\
\midrule
\multirow{11}{*}{\textbf{Barcelona}} 
 & Cocktail bar \& lounge & \multirow{6}{*}{Fall} & \multirow{11}{*}{Open} & \multirow{11}{*}{\textbf{\textsf{X}}} \\
 & Historic bakery & & & \\
 & Whiskey \& cocktail bar & & & \\
 & Mediterranean restaurant & & & \\
 & Record store & & & \\
 & Tapas bar & & & \\
\cline{2-3}
 & Cafe \& juice bar & \multirow{2}{*}{Rise \& Fall} & & \\
 & Record store & & & \\
\cline{2-3}
 & Comic store & \multirow{2}{*}{Steady} & & \\
 & Bubble tea shop & & & \\
\cline{2-3}
 & Tapas bar & Underdogs & & \\

\bottomrule
\end{tabularx}
\vspace{5pt}
\caption{Comparative qualitative summary of the venue profiles in the focal areas in Hackney (London) and Ciutat Vella (Barcelona).}
\label{tab:comparative_venues}
\end{table}

If Hackney represents a classic example of a maturely gentrified neighbourhood in Clay's four-stage model of gentrification \cite{clay1979neighborhood, Lees2025}, then the revitalization schemes in Ciutat Vella instead coincided with the 2008 economic recession, which incubated a speculative tourism gentrification \cite{CocolaGant2018, Domnguez2017} rather than a genuine improvement in the livelihood of local residents. The displacement of lower-income and older residents also means fewer regulars for local stores. This dynamic aligns with what \cite{Aalbers2018} characterizes as fifth-wave gentrification, in which neighbourhood change is driven less by the influx of middle-class residents and more by rental market speculation and housing financialization. 

Although the case study cannot directly quantify how much state policy or property-market speculation shapes a venue's trajectory, it explains the tension in our quantitative results. A neighbourhood can experience high income level and declining venue popularity simultaneously, and a medium-term, cross-sectional feature cannot by itself identify which stage of gentrification a neighbourhood is in. Urban transformation is a long-term, dynamic process, interacting with state policy, macroscopic economic forces, and street-level culture. More broadly, the case study serves as a cautionary lesson against asserting (or rejecting) any simple mechanisms that drive commercial spillover effects without considering the historical contexts of the space in focus.

\section*{Discussion}
Sustainable vibrancy and long-term success of urban venues, such as restaurants, parks, and department stores, are key indicators of a healthy urban economy. The popularity of urban venues has been a focal concern of urban data science, but the literature has largely approached it through the lens of failure (i.e., predicting closures). This framing treats success as the absence of death and reduces a venue's history to a binary outcome. Nonetheless, a venue that survives can flourish, stagnate, decline and recover, or climb steadily, and these paths are invisible to a survival analysis. To address this, building on existing work on POI success and urban vitality, this study followed venues over time, asking what shapes their popularity takes and how a venue's surroundings may play a role in determining its performance. 

We recovered six recurring trajectory shapes, consistent across London and Barcelona, suggesting these are signatures of how venues behave. The central result concerns what predicts them. Contrary to the expectation that a neighbourhood's socio-economic status and density determine a venue's performance, these static features carry little predictive power in our models. In fact, what predicts a venue's trajectory is the trajectories of its immediate neighbours. This spatial assortativity supports neighbourhood effects understood as a spillover mechanism \cite{glaeser2010agglomeration, durlauf2004neighbourhood, vanham2012neighbourhood, Daunfeldt2019}, and our results have demonstrated that spillover effects appear to emerge on the street level, but not the neighbourhood or ward level. Such a finding points at the possibility that large-scale static census data fail to predict performance because they average over the scale at which the effect is observable.

Despite the promising results, this work has several limitations. First, the Foursquare dataset is restricted to like counts in only two large European cities. This lack of cross-continent geographical comparison warrants further research to confirm the generalizability of the identified trajectory shapes. We also note that the socioeconomic status is operationalized differently across the two cities with two different datasets (i.e., IMD for London, household income for Barcelona), and they are measured at different spatial scales. Moreover, the assortativity radius is city-specific ($300$ m in London, $500$ m in Barcelona; see \hyperref[sec:assort]{Methods}). Our models and analyses are agnostic to the type of venues (e.g., restaurants, parks, grocery stores, pharmacies) that we are investigating, which potentially has intriguing interactions with their tendency to be (dis)assortative with its surrounding venues \cite{Yoshimura2020}. Additionally, although beyond the scope of this study, we encourage interested scholars to investigate whether these trajectories can also be observed when popularity is measured through alternative signals, such as actual visitation or comment counts, and whether such signals can inform decision-makers more accurately and reliably. It should also be acknowledged that, knowing that Large Language Models (LLMs) have grown to be a useful tool in urban science \cite{Fu2023, Fu2024, Rong2025}, we initially deployed LLM-based approaches also for the classification and prediction tasks. However, their performance fell short of the classical ML models, even under few-shot learning settings, and these results were therefore omitted from the final analysis. 

This work has uncovered patterns of venue success that is beyond a binary understanding of success versus failure. Vibrancy, on this account, is not a stable attribute of a neighbourhood but an emergent, local process. For policy-makers and business owners, because the spillover that governs a venue's trajectory operates street by street, tools intended to monitor or support local commerce are likely to be more informative when tailored to the immediate catchment than to the ward or borough. Our approach of combining large-scale behavioural trajectories with their spatial and historical context is readily available to other cities and signals, and offers a route to studying urban vitality as the dynamic, place-specific phenomenon.

\section*{Declarations}

\paragraph{Abbreviations.} DTW, Dynamic Time Warping; POI, Point of Interest; IMD, Index of Multiple Deprivation; MSE, Mean Squared Error; GRU, Gated Recurrent Unit; ETSC, Early time series classification; LLM, Large Language Model; KDE, Kernel Density Estimation; IQR, Interquartile Range; MAD, Mean Absolute Deviation; XGBoost, Extreme Gradient Boosting. 

\paragraph{Acknowledgement.} W.N.Y. acknowledges the support from Nokia Bell Labs during his research visit in 2025. R.D.C. was partially supported by COLINE DUT European Partnership project (F-DUT-2023-0037) and the Lagrange Project of the ISI Foundation, funded by Fondazione CRT.

\paragraph{Author contribution.} W.N.Y., L.M., D.Q. and R.D.C. conceived the study, developed the theoretical framework and designed the experiments. W.N.Y. performed all experiments, computation and analyses. L.M., D.Q. and R.D.C. supervised the project. W.N.Y. produced the figures. All authors discussed the results and contributed to the final manuscript.

\paragraph{Data availability.} The data used in this work is owned by Nokia Bell Labs and data access must be requested to and under agreement with Nokia Bell Labs.

\paragraph{Code availability.} All data processing and analyses are conducted with \verb|Python 3.12|. The code developed in this study will be made available at \url{https://github.com/ComplexConnectionsLab/urban_popularity_trajactory} upon publication.
\paragraph{Anonymization policy.} We anonymized specific POIs when being discussed. We do not identify any POIs with their names and only describe them generically.

\bibliography{references}

\newpage
\appendix
\setcounter{figure}{0}
\renewcommand{\thefigure}{S\arabic{figure}}
\renewcommand{\theHfigure}{S\arabic{figure}} 
\section*{Supplementary Material}

\section{Data description}\label{sec:supp_A1}
\begin{figure}[htp!]
    \centering
    \includegraphics[width=0.8\linewidth]{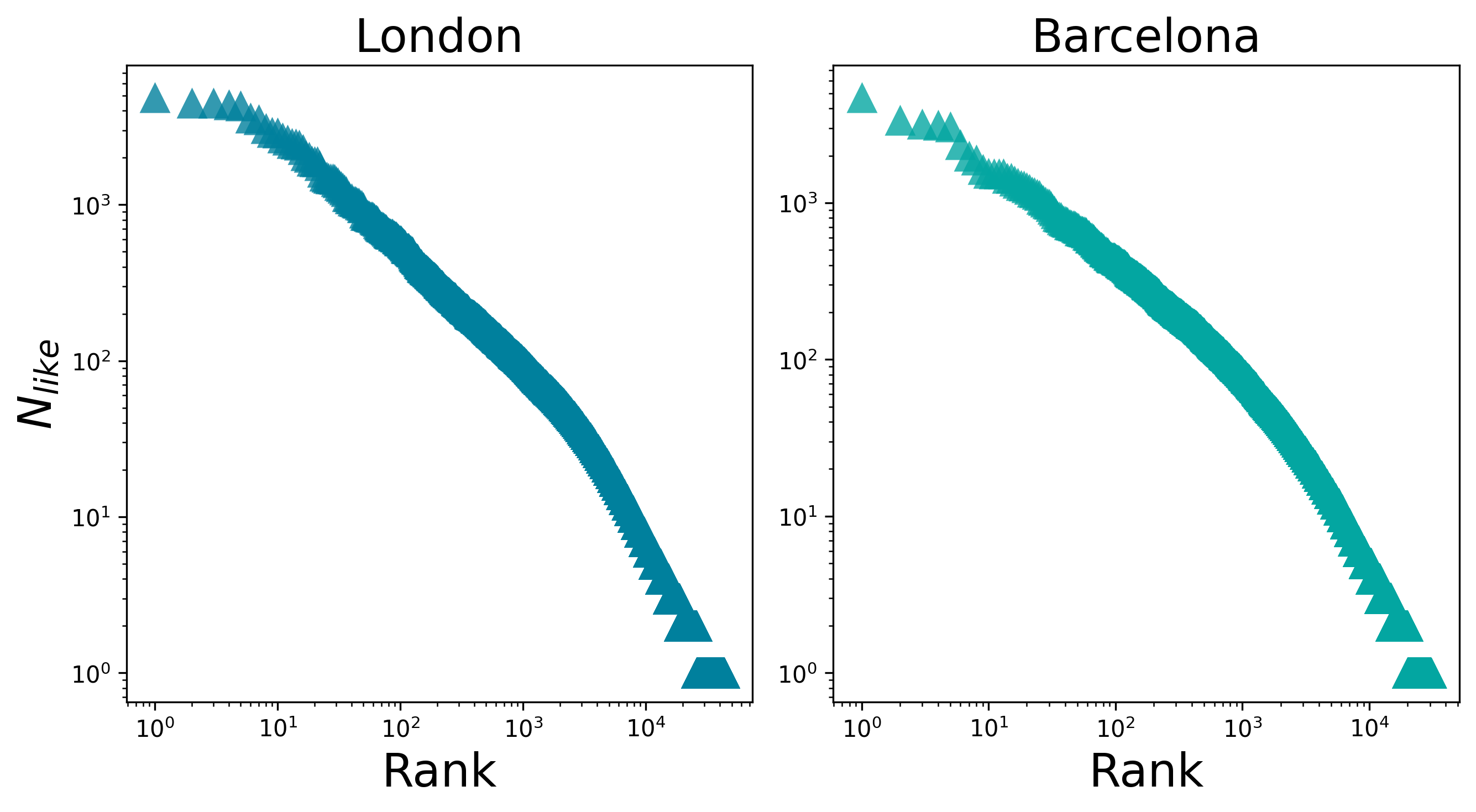}
    \caption{Log-Log Rank-frequency distribution of like counts in London and Barcelona. The power law fit yielded $\alpha \approx 2.34$ and $\alpha \approx 2.50$. However, we also note that, using the Vuong's closeness test, the distribution is not distinguishable from a lognormal distribution ($p > .05$) and thus we only describe this as a heavy-detailed distribution rather than a power law one.}
    \label{fig:supp_like_dist}
\end{figure}

\begin{figure}[htp!]
    \centering
    \includegraphics[width=0.8\linewidth]{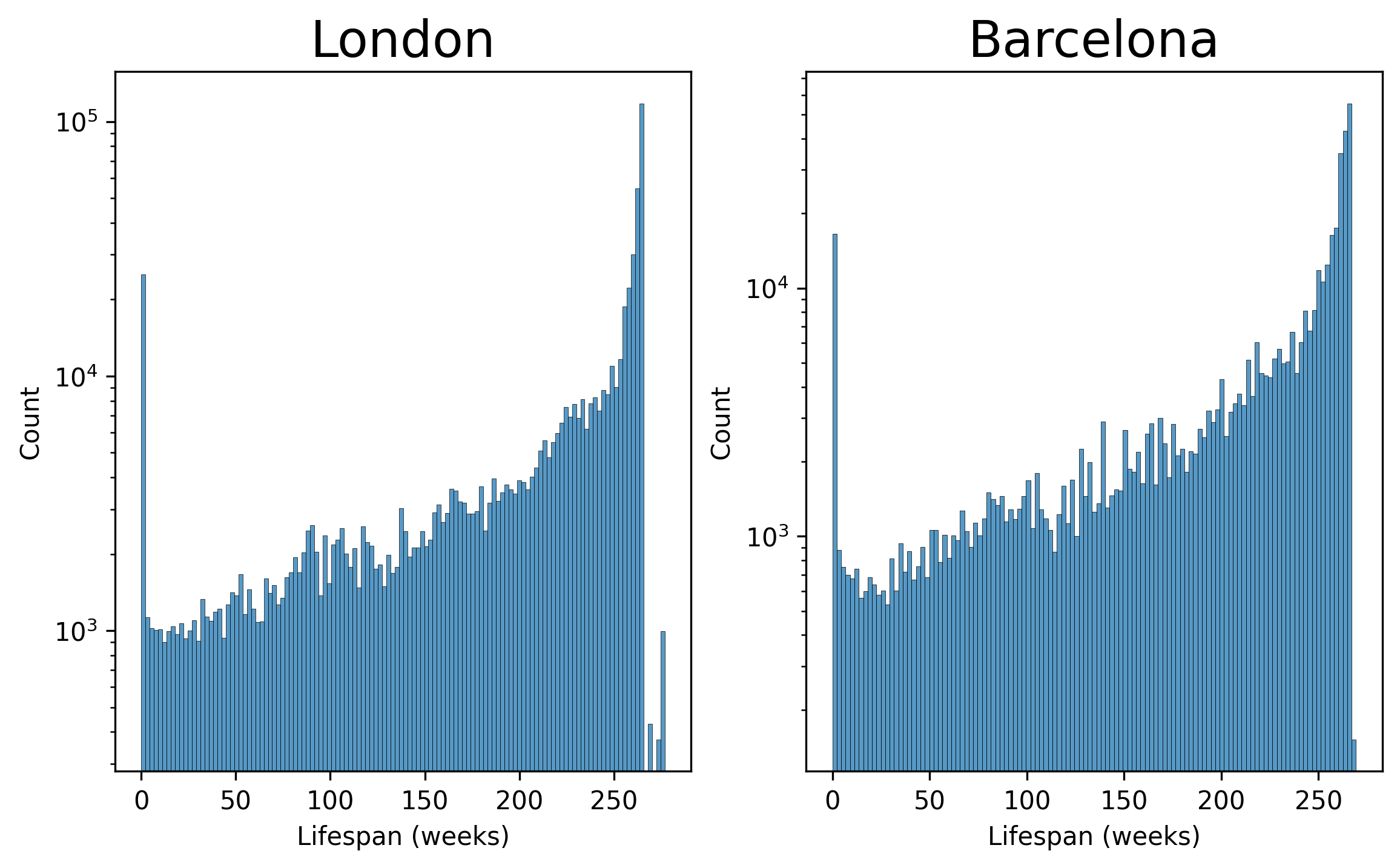}
    \caption{Log frequency distribution of lifespan (in weeks) in London and Barcelona. We estimated the lifespan by taking the day difference between the earliest and the latest like made for the venue.}
    \label{fig:supp_lifespan_dist}
\end{figure}

\begin{figure}[htp!]
    \centering
    \includegraphics[width=0.8\linewidth]{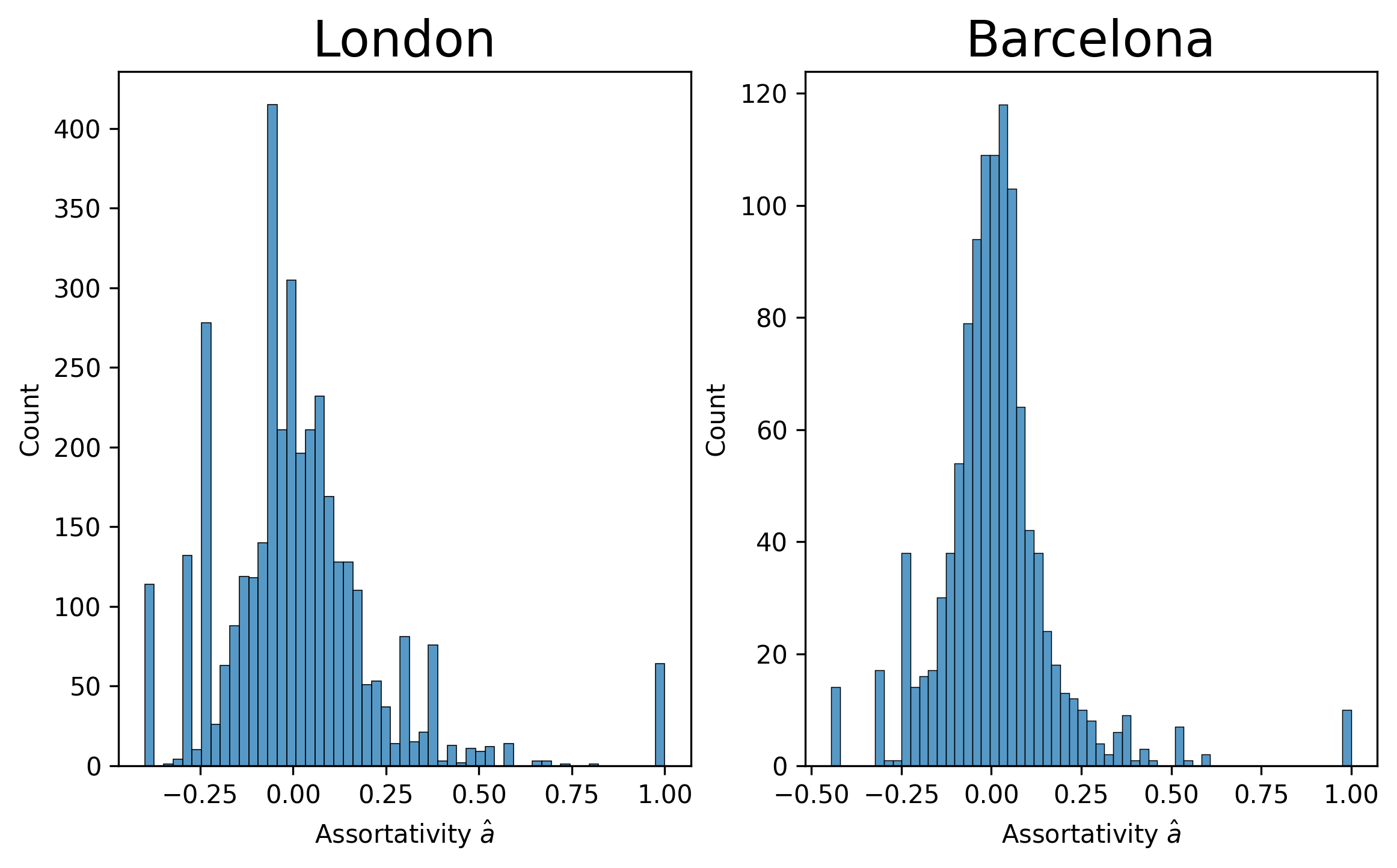}
    \caption{Frequency distribution of assortativity $\hat a$ in London and Barcelona. The distributions of assortativity coefficients for both cities are centred around zero. }
    \label{fig:supp_assort_dist}
\end{figure}

\begin{figure}[htp!]
    \centering
    \includegraphics[width=0.7\linewidth]{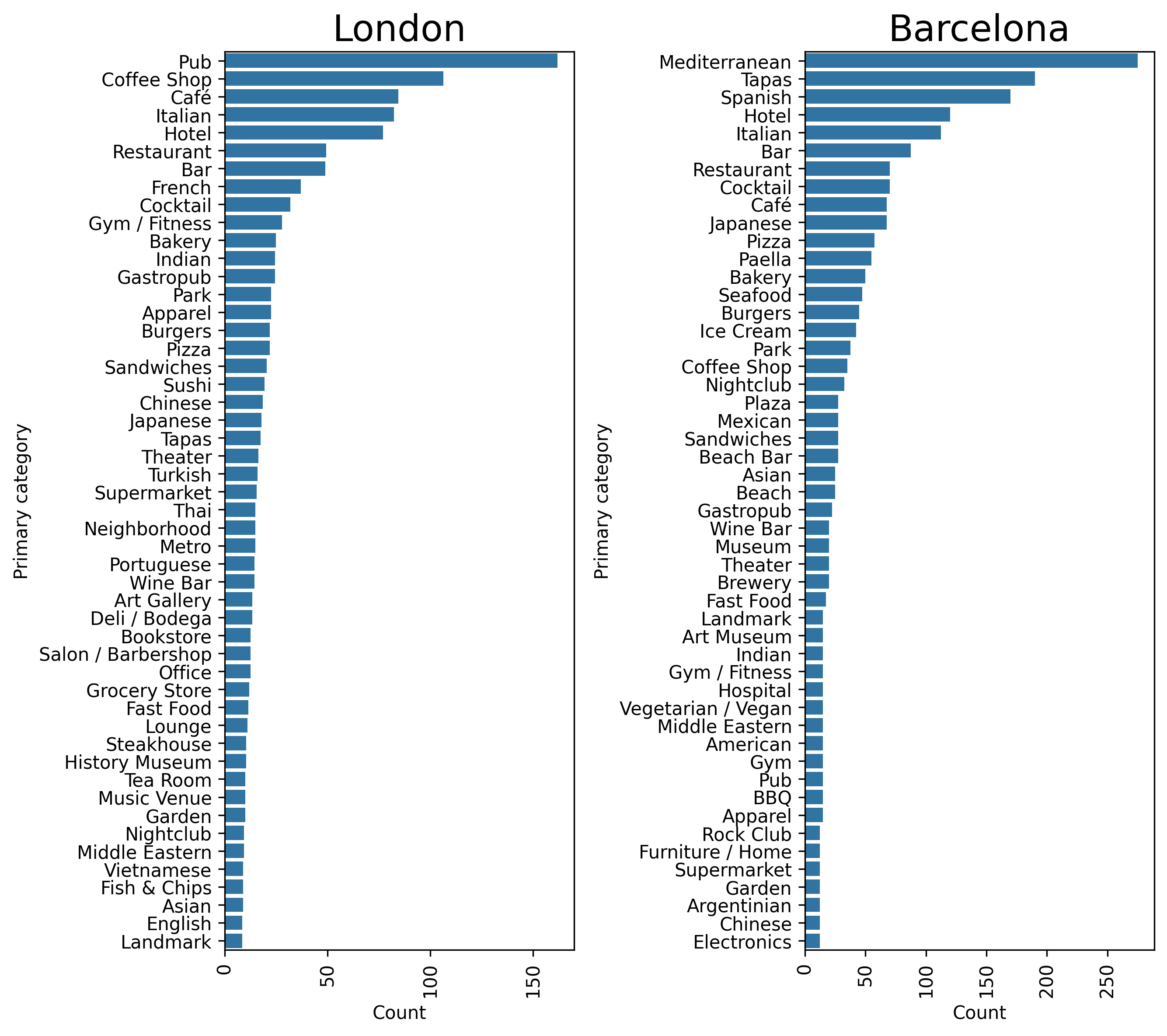}
    \caption{Frequency distribution of categories in London and Barcelona. Due to some venues having multiple categories, we have displayed here the first appearing tag as the primary category. The data show a culturally aligned distribution of popular category types, for example pubs and Mediterranean food venues being the most popular primary categories in London and Barcelona respectively.}
    \label{fig:supp_prim_count}
\end{figure}

\begin{figure}[htp!]
    \centering
    \includegraphics[width=0.7\linewidth]{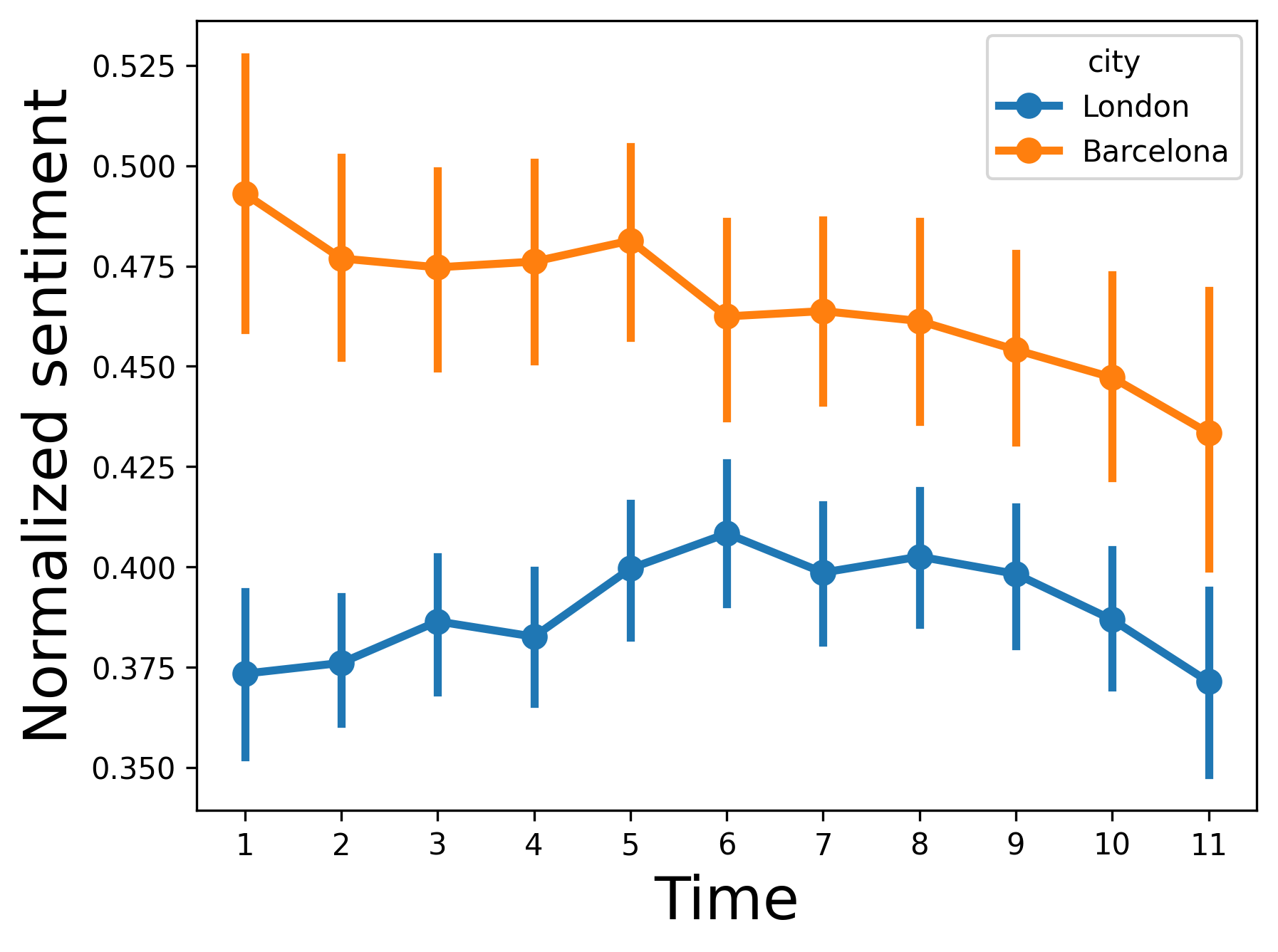}
    \caption{Normalized sentiment scores of venues in London (blue) and Barcelona (Orange) across 11 time points. Sentiment is first measured by DistilBERT-based multilingual sentiment model and rescaled to -2 to 2, aggregated on a six-monthly basis. Missing data are imputed by the mean of a given vector. Similar to the performance trajectory, we then length-normalized the sentiment time series to 11 points with a linear interpolation and extrapolation.}
    \label{fig:supp_sentiment_time}
\end{figure}

\newpage
\section{Description and data of cluster types}\label{sec:supp_clusters}

\begin{table}[htp!]
    \centering
    \renewcommand{\arraystretch}{1.4}
    \begin{tabular}{@{}>{\bfseries}l l@{}}
        \toprule
        \normalfont\textbf{Trajectory type} & \textbf{Description} \\
        \midrule
        Rise         & Popularity increases steadily over time. \\
        Fall         & Popularity decreases steadily over time. \\
        Rise \& Fall & Popularity rises at first, then declines in the later period. \\
        Steady       & Popularity is maintained over time without much fluctuation. \\
        Ephemeral    & Initially popular, but declines rapidly. \\
        Underdogs    & Initially unpopular, but gains popularity later on. \\
        \bottomrule
    \end{tabular}
    \vspace{5pt}
    \caption{The six popularity-trajectory types recovered by DTW clustering,
    common to both London and Barcelona.}
    \label{tab:trajectory-types}
\end{table}

\vspace{70pt}

\begin{figure}[htp!]
    \centering
    \includegraphics[width=0.5\linewidth]{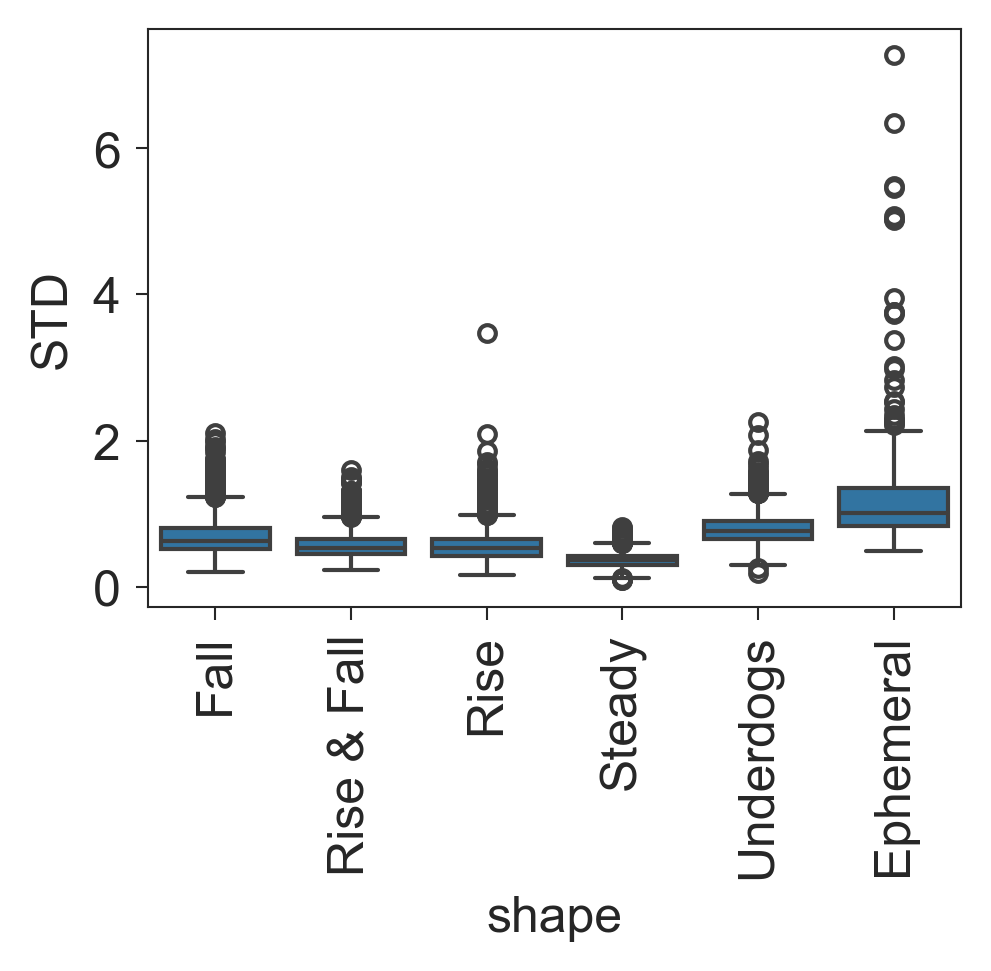}
    \caption{The standard deviation of the values in length-normalized trajectories for each shape. As Steady and Rise \& Fall are visually similar, we also fitted the trajectories to the quadratic form $ax^2 + bx +c$ and compared the average curvature of each group. A two-sided Mann-Whitney-U test shows significant differences betweeen the average curvature ($U = 3329523.0, p < .001$) with Rise \& Fall (average $a = 0.021$) trajectories being more curved than Steady ones (average $a = 0.016$).}
    \label{fig:supp_traj_shape_std}
\end{figure}

\newpage
\section{Additional results}
\subsection{Baseline trajectory type classification performance}
\begin{table*}[ht]
\centering
\begin{tabular}{@{}lcccccc c@{}}
\toprule
\textbf{Model} & \textbf{Fall} & \textbf{Rise} & \cellcolor{worst}\textbf{Rise \& Fall} & \textbf{Steady} & \cellcolor{best}\textbf{Ephemeral} & \textbf{Underdogs} & \textbf{Macro F1} \\
\midrule
Logistic Regression & 0.81 & 0.82 & \cellcolor{worst}0.51 & 0.72 & \cellcolor{best}0.85 & 0.66 & 0.729 \\
Random Forest & 0.78 & 0.82 & \cellcolor{worst}0.53 & 0.78 & \cellcolor{best}0.85 & 0.57 & 0.723 \\
XGBoost & \textbf{0.84} & \textbf{0.85} & \cellcolor{worst}\textbf{0.76} & \textbf{0.86} & \cellcolor{best}0.88 & \textbf{0.79} & \textbf{0.830} \\
\midrule
Class Average & 0.79 & 0.82 & \cellcolor{worst}0.54 & 0.77 & \cellcolor{best}0.86 & 0.56 & --- \\
\bottomrule
\end{tabular}
\vspace{2pt}
\caption{Per-class F1-scores across all models. Easiest class to predict highlighted in \colorbox{best}{green}, hardest in \colorbox{worst}{red}, and best results in \textbf{bold}.}
\label{tab:perclass_comparison}
\end{table*}

\subsection{Trajectory type classification performance with urban and venue features}\label{sec:SUPP_class}

\begin{table}[htp!]
\centering
\begin{tabular}{l l l c c}
\toprule
\textbf{City} & \textbf{Feature set} & \textbf{Model} & \textbf{Balanced Acc.} & \textbf{F1} \\
\midrule
\multirow{9}{*}{London}
& \multirow{3}{*}{Urban}
& Logistic Regression & 0.514 $\pm$ 0.017 & 0.513 $\pm$ 0.018 \\
& & Random Forest & 0.531 $\pm$ 0.009 & 0.529 $\pm$ 0.009 \\
& & XGBoost & \textbf{0.534 $\pm$ 0.008} & \textbf{0.533 $\pm$ 0.009} \\
\cmidrule(l){2-5}
& \multirow{3}{*}{Venue}
& Logistic Regression & 0.495 $\pm$ 0.012 & 0.491 $\pm$ 0.012 \\
& & Random Forest & \textbf{0.814 $\pm$ 0.019} & \textbf{0.813 $\pm$ 0.019} \\
& & XGBoost & 0.778 $\pm$ 0.018 & 0.778 $\pm$ 0.018 \\
\cmidrule(l){2-5}
& \multirow{3}{*}{Combined}
& Logistic Regression & 0.512 $\pm$ 0.007 & 0.511 $\pm$ 0.007 \\
& & Random Forest & 0.762 $\pm$ 0.006 & 0.762 $\pm$ 0.006 \\
& & XGBoost & \textbf{0.813 $\pm$ 0.012} & \textbf{0.812 $\pm$ 0.012} \\
\midrule
\multirow{9}{*}{Barcelona}
& \multirow{3}{*}{Urban}
& Logistic Regression & 0.545 $\pm$ 0.025 & 0.538 $\pm$ 0.023 \\
& & Random Forest & 0.541 $\pm$ 0.008 & 0.538 $\pm$ 0.006 \\
& & XGBoost & \textbf{0.560 $\pm$ 0.014} & \textbf{0.559 $\pm$ 0.014} \\
\cmidrule(l){2-5}
& \multirow{3}{*}{Venue}
& Logistic Regression & 0.529 $\pm$ 0.023 & 0.523 $\pm$ 0.021 \\
& & Random Forest & \textbf{0.688 $\pm$ 0.023} & \textbf{0.688 $\pm$ 0.024} \\
& & XGBoost & 0.674 $\pm$ 0.020 & 0.673 $\pm$ 0.020 \\
\cmidrule(l){2-5}
& \multirow{3}{*}{Combined}
& Logistic Regression & 0.543 $\pm$ 0.033 & 0.542 $\pm$ 0.033 \\
& & Random Forest & 0.715 $\pm$ 0.020 & 0.714 $\pm$ 0.019 \\
& & XGBoost & \textbf{0.745 $\pm$ 0.026} & \textbf{0.745 $\pm$ 0.027} \\
\bottomrule
\end{tabular}
\vspace{5pt}
\caption{Binary classification performance (mean $\pm$ std) with 5-fold cross-validation by city, feature set, and model. Best model per feature-set block shown in \textbf{bold}. Random seed is $98765$.}
\label{tab:binary_results}
\end{table}
\newpage
\begin{table}[htp!]
\centering
\begin{tabular}{l l l c c}
\toprule
\textbf{City} & \textbf{Feature set} & \textbf{Model} & \textbf{Balanced Acc.} & \textbf{F1} \\
\midrule
\multirow{9}{*}{London}
 & \multirow{3}{*}{Urban}
   & Logistic Regression & 0.158 \(\pm\) 0.017 & 0.100 \(\pm\) 0.030 \\
 & & Random Forest        & 0.175 \(\pm\) 0.011 & 0.160 \(\pm\) 0.006 \\
 & & XGBoost              & \textbf{0.175 \(\pm\) 0.013} & \textbf{0.166 \(\pm\) 0.012} \\
\cmidrule(l){2-5}
 & \multirow{3}{*}{Venue}
   & Logistic Regression & 0.191 \(\pm\) 0.007 & 0.112 \(\pm\) 0.016 \\
 & & Random Forest        & \textbf{0.694 \(\pm\) 0.031} & \textbf{0.686 \(\pm\) 0.028} \\
 & & XGBoost              & 0.620 \(\pm\) 0.030 & 0.637 \(\pm\) 0.030 \\
\cmidrule(l){2-5}
 & \multirow{3}{*}{Combined}
   & Logistic Regression & 0.183 \(\pm\) 0.032 & 0.132 \(\pm\) 0.026 \\
 & & Random Forest        & 0.624 \(\pm\) 0.008 & 0.612 \(\pm\) 0.008 \\
 & & XGBoost              & \textbf{0.668 \(\pm\) 0.015} & \textbf{0.688 \(\pm\) 0.016} \\
\midrule
\multirow{9}{*}{Barcelona}
 & \multirow{3}{*}{Urban}
   & Logistic Regression & 0.191 \(\pm\) 0.023 & 0.119 \(\pm\) 0.014 \\
 & & Random Forest        & 0.171 \(\pm\) 0.018 & 0.159 \(\pm\) 0.015 \\
 & & XGBoost              & \textbf{0.198 \(\pm\) 0.023} & \textbf{0.199 \(\pm\) 0.037} \\
\cmidrule(l){2-5}
 & \multirow{3}{*}{Venue}
   & Logistic Regression & 0.221 \(\pm\) 0.025 & 0.131 \(\pm\) 0.023 \\
 & & Random Forest        & 0.488 \(\pm\) 0.045 & 0.472 \(\pm\) 0.035 \\
 & & XGBoost              & \textbf{0.527 \(\pm\) 0.037} & \textbf{0.535 \(\pm\) 0.031} \\
\cmidrule(l){2-5}
 & \multirow{3}{*}{Combined}
   & Logistic Regression & 0.226 \(\pm\) 0.032 & 0.166 \(\pm\) 0.023 \\
 & & Random Forest        & \textbf{0.578 \(\pm\) 0.042} & 0.539 \(\pm\) 0.043 \\
 & & XGBoost              & 0.557 \(\pm\) 0.056 & \textbf{0.575 \(\pm\) 0.057} \\
\bottomrule
\end{tabular}
\vspace{5pt}
\caption{Multiclass classification performance (mean $\pm$ std) with 5-fold cross-validation by city, feature set, and model. Best model per feature-set block shown in \textbf{bold}. Random seed is $98765$.}
\label{tab:multiclass_results}
\end{table}

\newpage
\begin{table}[h!]
\centering
\begin{tabular}{l l c c c}
\toprule
\textbf{Feature set} & \textbf{Model} & \textbf{F1 (London)} & \textbf{F1 (Barcelona)} & \textbf{$\Delta$F1} \\
\midrule
\multicolumn{5}{l}{\textit{Binary}} \\
\midrule
Urban    & Logistic Regression & 0.513 & 0.538 & $-$0.025 \\
Urban    & Random Forest       & 0.529 & 0.538 & $-$0.009 \\
Urban    & XGBoost             & 0.533 & 0.559 & $-$0.026 \\
Venue    & Logistic Regression & 0.491 & 0.523 & $-$0.031 \\
Venue    & Random Forest       & 0.813 & 0.688 & $+$0.126 \\
Venue    & XGBoost             & 0.778 & 0.673 & $+$0.105 \\
Combined & Logistic Regression & 0.511 & 0.542 & $-$0.032 \\
Combined & Random Forest       & 0.762 & 0.714 & $+$0.048 \\
Combined & XGBoost             & 0.812 & 0.745 & $+$0.068 \\
\midrule
\multicolumn{3}{l}{Mean F1} & 0.638 & 0.613 \\
\midrule
\multicolumn{5}{l}{\textit{Multiclass}} \\
\midrule
Urban    & Logistic Regression & 0.100 & 0.119 & $-$0.018 \\
Urban    & Random Forest       & 0.160 & 0.159 & $+$0.001 \\
Urban    & XGBoost             & 0.166 & 0.199 & $-$0.032 \\
Venue    & Logistic Regression & 0.112 & 0.131 & $-$0.018 \\
Venue    & Random Forest       & 0.686 & 0.472 & $+$0.214 \\
Venue    & XGBoost             & 0.637 & 0.535 & $+$0.102 \\
Combined & Logistic Regression & 0.132 & 0.166 & $-$0.034 \\
Combined & Random Forest       & 0.612 & 0.539 & $+$0.073 \\
Combined & XGBoost             & 0.688 & 0.575 & $+$0.113 \\
\midrule
\multicolumn{3}{l}{Mean F1} & 0.366 & 0.322 \\
\bottomrule
\end{tabular}
\vspace{5pt}
\caption{City comparison of F1 scores (averaged across folds) across feature sets and models, binary and multiclass tasks. $\Delta$F1 = F1(London) $-$ F1(Barcelona). Random seed is $98765$.}
\label{tab:city_comparison}
\end{table}

\newpage
\subsection{Feature importance in type classification using urban and venue data}
\begin{table}[hp!]
\centering

\vspace{6pt}

\textbf{(a) Logistic Regression}
\vspace{4pt}

\begin{tabular}{llcccc}
\toprule
\textbf{City} & \textbf{Task} & \textbf{Deprivation} & \textbf{Pop.\ Density} & \textbf{Uniqueness} & \textbf{Assortativity} \\
\midrule
London    & Binary     & $.011 \pm .004$  & $.025 \pm .008$  & $.001 \pm .003$ & $.011 \pm .005$ \\
          & Multiclass & $.010 \pm .003$  & $.010 \pm .005$  & $.006 \pm .004$ & $.014 \pm .004$ \\
\midrule
Barcelona & Binary     & $-.003 \pm .002$ & $\mathbf{.041 \pm .014}$ & $.007 \pm .008$ & $-.005 \pm .006$ \\
          & Multiclass & $.003 \pm .006$  & $\mathbf{.021 \pm .007}$ & $.010 \pm .005$ & $.021 \pm .005$ \\
\bottomrule
\end{tabular}

\vspace{12pt}

\textbf{(b) Random Forest}
\vspace{4pt}

\begin{tabular}{llcccc}
\toprule
\textbf{City} & \textbf{Task} & \textbf{Deprivation} & \textbf{Pop.\ Density} & \textbf{Uniqueness} & \textbf{Assortativity} \\
\midrule
London    & Binary     & $0.179 \pm 0.005$ & $0.262 \pm 0.006$ & $0.076 \pm 0.003$ & $\mathbf{0.427 \pm 0.007}$ \\
          & Multiclass & $0.291 \pm 0.007$ & $0.359 \pm 0.008$ & $0.156 \pm 0.005$ & $\mathbf{0.753 \pm 0.008}$ \\
\midrule
Barcelona & Binary     & $0.255 \pm 0.012$ & $0.206 \pm 0.009$ & $0.106 \pm 0.007$ & $\mathbf{0.390 \pm 0.014}$ \\
          & Multiclass & $0.342 \pm 0.013$ & $0.306 \pm 0.016$ & $0.162 \pm 0.010$ & $\mathbf{0.722 \pm 0.012}$ \\
\bottomrule
\end{tabular}

\vspace{12pt}

\textbf{(c) XGBoost}
\vspace{4pt}

\begin{tabular}{llcccc}
\toprule
\textbf{City} & \textbf{Task} & \textbf{Deprivation} & \textbf{Pop.\ Density} & \textbf{Uniqueness} & \textbf{Assortativity} \\
\midrule
London    & Binary     & $0.081 \pm 0.003$ & $0.113 \pm 0.005$ & $0.028 \pm 0.002$ & $\mathbf{0.421 \pm 0.007}$ \\
          & Multiclass & $0.168 \pm 0.005$ & $0.204 \pm 0.006$ & $0.078 \pm 0.003$ & $\mathbf{0.765 \pm 0.006}$ \\
\midrule
Barcelona & Binary     & $0.172 \pm 0.008$ & $0.160 \pm 0.010$ & $0.082 \pm 0.006$ & $\mathbf{0.400 \pm 0.010}$ \\
          & Multiclass & $0.267 \pm 0.012$ & $0.218 \pm 0.011$ & $0.127 \pm 0.009$ & $\mathbf{0.748 \pm 0.015}$ \\
\bottomrule
\end{tabular}
\vspace{10pt}
\caption{Permutation importance (mean $\pm$ std) for predicting venue popularity trend using 5-fold cross-validation. Bolded cells indicate the most important feature for each model: \textbf{(a)}~Logistic Regression, \textbf{(b)}~Random Forest, \textbf{(c)}~XGBoost. Random seed is $98765$.}
\label{tab:feature_import}

\end{table}

\newpage
\subsection{Baseline performance and ablation test for early classification task}

\begin{figure}[htp!]
    \centering
    \includegraphics[width=0.9\linewidth]{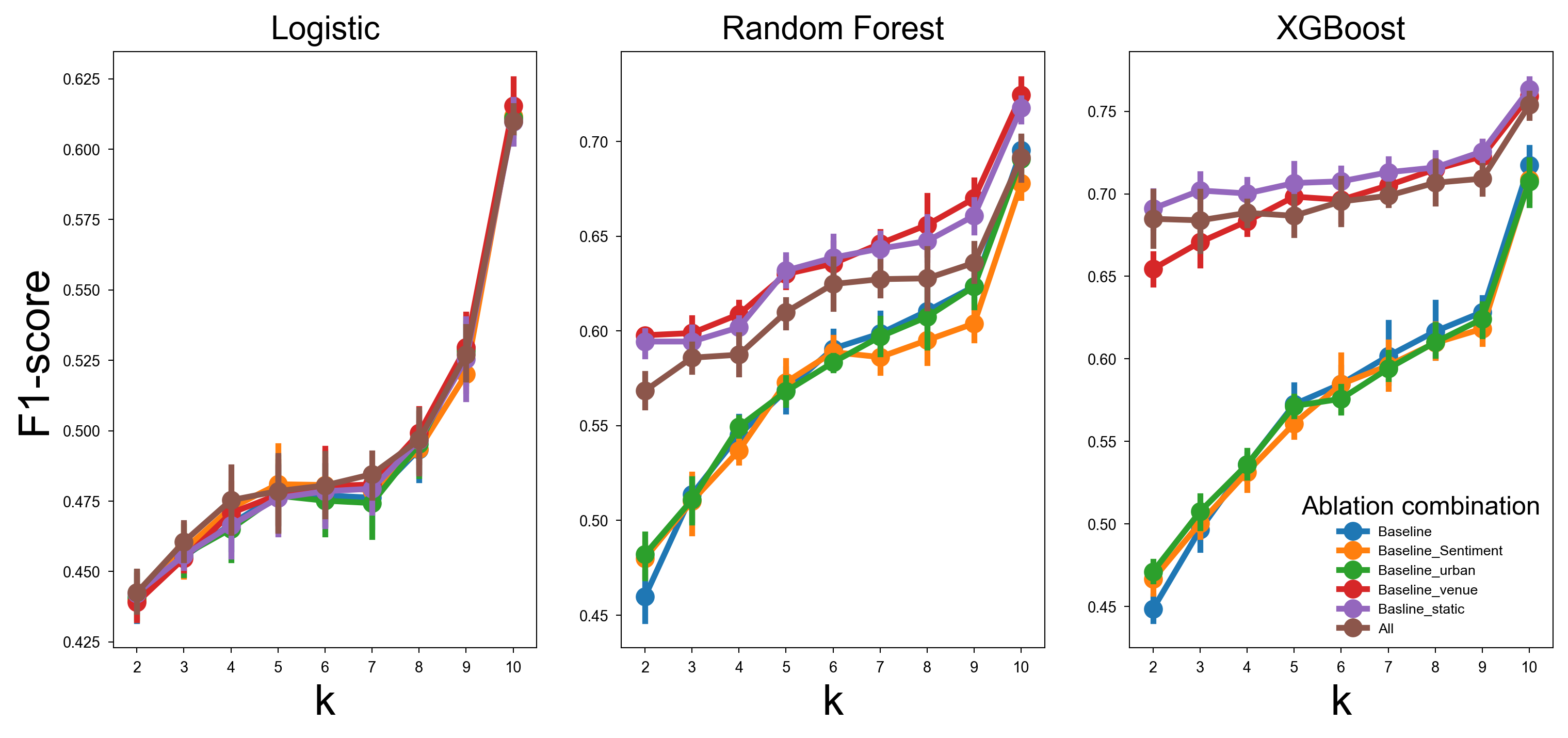}
    \caption{Macro F1-score as a function of early-classification horizon $k$, across feature ablations. Error bars denote $\pm$ std across 5 folds. \texttt{Baseline\_venue}, \texttt{Baseline\_static} (i.e., urban + venue), and \texttt{All} consistently outperform \texttt{Baseline}, \texttt{Baseline\_Sentiment}, and \texttt{Baseline\_urban} for $k \leq 9$. All ablations exhibit a sharp increase in F1-score as $k$ approaches $10$, consistent with the sequence itself becoming sufficiently informative once most of it has been observed.}
\label{fig:early_class_ablation_by_k}
\end{figure}



\subsection{Spatial null model results with varied $R$}
\begin{table}[htbp]
\centering

\begin{tabular}{l c c c c r}
\toprule
\textbf{City} & \textbf{$R$} & \textbf{Obs.\ variance} & \textbf{Null variance} & \textbf{$p$} & \textbf{\# Circles} \\
\midrule
\multirow{4}{*}{London}
 & 200 & 0.00805 & 0.00740 & 0.282        & 347 \\
 & 250 & 0.00687 & 0.00597 & 0.218        & 384 \\
 & 300 & 0.00664 & 0.00491 & 0.080  & 409 \\
 & 500 & 0.00292 & 0.00271 & 0.317        & 468 \\
\midrule
\multirow{4}{*}{Barcelona}
 & 200 & 0.01141 & 0.00666 & 0.022$^{*}$ & 327 \\
 & 250 & 0.01025 & 0.00641 & 0.060  & 375 \\
 & 300 & 0.00747 & 0.00538 & 0.099  & 403 \\
 & 500 & 0.00305 & 0.00242 & 0.158        & 471 \\
\bottomrule
\end{tabular}
\vspace{4pt}
\caption{Spatial null model for $\hat \alpha$ in London and Barcelona. * indicates a significant difference between the observed and null variance of $\hat \alpha$.}
\label{tab:random_circle_test}
\end{table}

\newpage

\end{document}